\documentclass[12pt]{article}
\usepackage{hyperref}
\usepackage{amssymb,amsmath,amsfonts,mathtools}
\usepackage{commath}
\usepackage{epsfig}
\usepackage{enumerate}
\usepackage{subfigure}
\usepackage[T1]{fontenc}

\usepackage[bulletsep]{collref}

\def\clock{{\count0=\time
           \divide\count0 60
           \ifnum\count0<10 0\fi\the\count0
           \multiply\count0 -60 \advance\count0 \time
           :\ifnum\count0<10 0\fi \the\count0
         }}
\newcommand{\timestamp}{{\small\vbox{\hbox{\tt\jobname.tex}
\hbox{\the\day/\the\month/\the\year, \clock}}}}

\newcommand{\CC}{\mathcal{C}}

\newcommand{\CP}{\mathcal{P}}

\newcommand{\C}{\mathbb{C}}
\newcommand{\R}{\mathbb{R}}

\newcommand{\nn}{\nonumber}

\DeclareMathOperator{\arcsinh}{arcsinh}

\newcommand{\ads}{\textup{\textrm{AdS}}}
\newcommand{\adsBH}[1]{\ensuremath{\ads_{#1}\textrm{-Schwarzschild}}}

\newcommand{\sphere}{\textup{\textrm{S}}}

\newcommand{\order}{\ensuremath{O}}

\makeatletter
\let\old@startsection=\@startsection
\let\oldl@section=\l@section
\renewcommand{\@startsection}[6]{\old@startsection{#1}{#2}{#3}{#4}{#5}{#6\mathversion{bold}}}
\renewcommand{\l@section}[2]{\oldl@section{\mathversion{bold}#1}{#2}}
\makeatother

\numberwithin{equation}{section}

\def\[{\begin{equation}}
\def\]{\end{equation}}
\newcommand{\be}{\begin{eqnarray}}
\newcommand{\ee}{\end{eqnarray}}

\newcommand{\grp}[1]{\mathrm{#1}}

\newcommand{\grSU}{\grp{SU}}
\newcommand{\grU}{\grp{U}}

\newcommand{\half}{\frac{1}{2}}
\newcommand{\p}{\partial}

\newcommand{\bho}{\bar{\rho}}

\newcommand{\mubh}{\ensuremath{\mu_{b,\rm h}}}

\newcommand{\Pbc}{\ensuremath{\bar{\mathcal{P}}}}

\usepackage{color}

\def\up{\upsilon}

\def\R{ {\mathbb R} }
\def\C{ {\mathbb C} }

\begin{document}
\renewcommand{\thefootnote}{\arabic{footnote}}
 
\overfullrule=0pt
\parskip=2pt
\parindent=12pt
\headheight=0in \headsep=0in \topmargin=0in \oddsidemargin=0in

\vspace{ -3cm} \thispagestyle{empty} \vspace{-1cm}
\begin{flushright} 
\footnotesize
\end{flushright}%

\begin{center}
\vspace{1.2cm}
{\Large\bf \mathversion{bold}
Monopole correlation functions and holographic phases of matter in
2+1 dimensions }

 \vspace{0.8cm} {
  T.~Alho$^{a,}$\footnote{ {\tt alho@hi.is}},
 V.~Giangreco~M. Puletti$^{a,}$\footnote{ {\tt vgmp@hi.is}},
  R.~Pourhasan$^{a,}$\footnote{ {\tt razieh@hi.is}}, 
L.~Thorlacius$^{a,b, }$\footnote{ {\tt lth@hi.is}},
}
 \vskip  0.5cm

\small
{\em
$^{a}$University of Iceland,
Science Institute,
Dunhaga 3,  107 Reykjavik, Iceland
\vskip 0.05cm
$^{b}$ The Oskar Klein Centre for Cosmoparticle Physics, Department of Physics, 
Stockholm University, AlbaNova University Centre, 10691 Stockholm, Sweden
}
\normalsize

 \end{center}

\vspace{0.3cm}
\begin{abstract}

The strong coupling dynamics of a 2+1 dimensional U(1) gauge theory 
coupled to charged matter is holographically modeled via a top-down 
construction with intersecting D3- and D5-branes. We explore the resulting 
phase diagram at finite temperature and charge density using correlation 
functions of monopole operators, dual to magnetically charged particles 
in the higher-dimensional bulk theory, as a diagnostic.

\end{abstract}
\newpage

\tableofcontents

\setcounter{footnote}{0}
\newpage

\section{Introduction}
\label{sec:Introduction}

Gauge/gravity duality provides an interesting setting for the study of compressible
quantum phases, where strongly correlated quantum dynamics is encoded into 
spacetime geometry in a gravitational dual description.
The best understood cases all involve supersymmetric Yang-Mills 
theories in a large $N$ limit, which are rather exotic from the point of 
view of many-body physics. Emergent gauge fields 
are known to arise in various quantum critical systems but these are 
almost exclusively $\grU(1)$ fields and do not immediately 
lend themselves to a large $N$ treatment. Gauge/gravity duality does,
however, offer a rare glimpse into strongly coupled dynamics in a setting 
where explicit computations are relatively straightforward and 
some aspects of the dynamics, in particular at finite temperature 
and density, may be generic to more general strongly coupled field 
theories.

Motivated by recent work of N.\ Iqbal \cite{Iqbal:2014cga}, we apply the formalism of 
gauge/gravity duality to map out the phase diagram of a 2+1-dimensional many body
system with a conserved $\grU(1)$ current at finite temperature and charge density.
We use correlation functions of suitably defined magnetic monopole operators 
to probe the  relevant physics \cite{Iqbal:2014cga,Faulkner:2012gt}. The fact that magnetic
monopoles can strongly influence the infrared behavior of gauge theories is well known.
For instance, the key role of monopoles in precipitating confinement in 2+1~dimensional 
gauge dynamics was emphasized in the pioneering work of 
Polyakov \cite{Polyakov:1975rs,Polyakov:1976fu}. 
In a condensed matter context, monopoles provide an order parameter for the transition from 
anti-ferromagnetic order to valence bond solid in a gauge theory description of certain 
two-dimensional lattice anti-ferromagnets \cite{Read:1990zza, Read:1989aa, Murthy:1989ps}.
The phase transition is continuous and described by a $\C P^N$ model with monopoles
condensing at the critical point, which has motivated the computation of monopole 
correlation functions in the $\C P^N$ model in a $1/N$ 
expansion \cite{Pufu:2013vpa,Pufu:2013eda,Dyer:2015zha}.\footnote{See \cite{Chester:2015wao} 
for a study of monopole operators by means of $4-\epsilon$ expansion.}

In a 2+1-dimensional gauge theory, a magnetic monopole operator, $\mathcal M(x_m)$, 
corresponds to a localized defect where a magnetic flux is inserted. Such operators belong 
to a more general class of topological disorder operators \cite{Borokhov:2002ib}. 
Their construction in terms of singular boundary conditions in a path integral formalism is
outlined in \cite{Iqbal:2014cga}. Due to flux quantization, monopole operators are 
intrinsically non-perturbative and difficult to handle using conventional field theory techniques.
In holography, on the other hand, correlation functions of monopole operators have a 
straightforward geometric representation and can be numerically evaluated using relatively 
simple methods. 

The holographic description of magnetic monopole operators, in terms of intersecting D-branes, 
that we will be using was developed in \cite{Iqbal:2014cga,Filev:2014mwa}.%
\footnote{For other 
works related to holographic monopoles, see {\it e.g.} \cite{Bolognesi:2010nb, Sutcliffe:2011sr, Rougemont:2015gia}.} 
The starting point for the construction is a well-known top-down model for a
2+1-dimensional field theory living on the intersection of  a single D5-brane and a large 
number $N$ of coincident 
D3-branes \cite{DeWolfe:2001bt, Erdmenger:2002ed, Karch:2002sh, Skenderis:2002vf}.
In this model, the D5-brane is treated as a probe brane in the $\ads_5\times\sphere^5$ 
background geometry sourced by the D3-branes. 
The embedding of the D5-brane into the D3-brane geometry is obtained by minimizing 
the DBI action of the D5-brane in an $\ads_5\times\sphere^5$ background (we 
review the calculation in Section~\ref{sec:preliminaries}).
There exists a solution where the D5-brane wraps an S$_2$ of fixed radius inside
the S$_5$ and extends along an $\ads_4$ subspace of the $\ads_5$. This corresponds 
to a conformally invariant state in the dual 2+1-dimensional boundary theory. There are
other solutions where the D5-brane embedding caps off at a finite radial coordinate,
corresponding to a deformation away from criticality and a mass gap 
in the 2+1-dimensional theory. 

Open strings stretching between the D3- and D5-branes give rise to matter fields in the 
fundamental representation of the SU($N$) gauge group that are localised on the 
2+1-dimensional intersection. The boundary theory also has a conserved global 
$\grU(1)$ current, which corresponds under gauge/gravity duality to a bulk $\grU(1)$ 
gauge field living in $\ads_4$. In general, a monopole operator inserted at the 
2+1-dimensional boundary corresponds to a bulk field carrying magnetic charge under 
the bulk  gauge field \cite{Sachdev:2012ks,Witten:2003ya}. In the top-down construction
of \cite{Iqbal:2014cga} the monopoles are realized as a probe D3-brane, 
oriented in such a way as to appear as a one-dimensional curve in $\ads_4$, 
with the remaining world-volume coordinates filling (at most) half an $\sphere^3$ 
in $\sphere^5$ and ending on the $\sphere^2$ wrapped by the D5-brane.
A D$(p-2)$-brane ending on a D$p$-brane carries magnetic charge in the D$p$ world-volume \cite{Strominger:1995ac} and thus the probe D3-brane represents a magnetically charged 
particle in $\ads_4$.  If the D3 curve reaches the $\ads_4$ boundary at a point $x_m$, 
it corresponds to an insertion of a magnetic flux at that point, {\it i.e.} a boundary 
monopole operator. We review the construction in more detail in 
Section~\ref{sec:monopole-nocharge} and extend it to finite temperature backgrounds.

In the large $N$ limit, the two-point function of boundary monopole operators is given
by the on-shell D3 action,
\be\label{monopole-corr}
\langle \mathcal M( \Delta x) \mathcal M^\dagger (0)\rangle \sim e^{- S_{D3}[\Delta x]}\,.
\ee
The D3-brane action consists of the usual DBI term and a magnetic coupling term. 
The DBI term is proportional to the length of the curve in $\ads_4$ traced out by the 
D3-brane, in a metric that 
depends on the D5-brane embedding, while the remaining
term involves the integral of the magnetic dual of the world-volume gauge field along the 
same curve. The magnetic coupling will play a key role when we 
consider backgrounds at finite charge density.

In a charge gapped phase monopoles are expected to condense at large enough separation, 
that is their equal-time two-point function is expected to saturate with distance between the 
monopole insertion points \cite{Sachdev:2012ks}. 
In a Fermi-liquid phase (a compressible phase with non-zero charge density and no broken symmetries), field theory computations become rather involved due to the non-perturbative 
nature of the monopoles but \cite{Lee:2008cl} predicted a faster 
than power-law falloff for the monopole equal-time two-point function.  
This behaviour was indeed found in the holographic computation in~\cite{Iqbal:2014cga}, 
which gave a constant value for the monopole correlation as 
a function of distance in a charge gapped case, and a Gaussian falloff at large 
separation in a compressible phase.

Our goal is to understand how turning on a non-zero temperature affects monopole 
correlation functions. In particular, we wish to determine whether the spatial dependence 
of the monopole equal-time two-point function can still serve as an order parameter 
for phase transitions at finite $T$.  
Our starting point is the holographic model employed in \cite{Iqbal:2014cga}, except now 
the D3-brane background is an $\ads_5$-Schwarzschild$\times\sphere^5$ black brane,
and we investigate the behavior of monopole correlation functions across the rather
rich phase diagram spanned by temperature and charge density. Similar questions 
can in principle be addressed in other holographic models, including various 
phenomenologically motivated bottom-up models. It would be interesting to 
pursue this in future work but for now we will take advantage of the higher-dimensional 
geometric perspective provided by the specific top-down construction of \cite{Iqbal:2014cga}.

The paper is organized as follows. 
In Section \ref{sec:zero-density} we review the D3/D5-brane construction at finite
temperature. We then introduce a monopole D3-brane and compute the D3-brane action
that gives the monopole two-point function.
In Section~\ref{sec:finite_density} we turn to the Fermi-liquid phase at finite charge density. 
After briefly introducing the relevant background D3/D5-brane solutions, we proceed to the 
D3-brane action, and map out the corresponding phase diagram. We conclude with a brief 
discussion in Section~\ref{sec:discussions}. 
Our conventions and definitions of the action functionals governing the probe D-brane 
dynamics studied in the paper are collected in Appendix~\ref{app:convention}. This is
followed in Appendix~\ref{app:thermodynamics-D5} by a short discussion of the boundary 
counter-terms that are required for the regularisation of the D5-brane free energy. 
A detailed examination of the on-shell D3-brane action at finite charge density and 
temperature is carried out in Appendix~\ref{sec:D3_general} and referred to in the main 
text. In Appendix~\ref{asymptotia} we consider asymptotic limits of model parameters, 
where analytic results can be obtained. This complements the numerical investigation 
in the rest of the paper and provides a useful check on the numerics.

\section{Monopole correlators at finite temperature}
\label{sec:zero-density}

\subsection{Probe D5-brane in a black 3-brane background}
\label{sec:preliminaries}

Throughout the paper we consider probe D-branes in the finite temperature near-horizon 
geometry of $N$ D3-branes, 
\be\label{bgmetricu}
ds^2 =\frac{u^2}{L^2}\Big[-h(u)\,dt^2{+}dx^2{+}dy^2{+}dx_{\perp}^2\Big]
{+}\frac{L^2}{u^2}\Big[\frac{du^2}{h(u)}
{+}u^2(d\psi^2{+}\sin^2\psi\, d\Omega_2^2 {+}\cos^2\psi \, d\tilde{\Omega}_2^2)\Big], \nn\\
\ee
where $h(u)=1-(u_0/u)^4$, $d\Omega_2^2 = d\theta^2+\sin^2\theta d\phi^2$, 
$d\tilde{\Omega}_2^2 = d\tilde{\theta}^2+\sin^2\tilde{\theta} d\tilde{\phi}^2$, and $L$ is a characteristic length scale. In these coordinates there
is an event horizon at $u=u_0$ and the asymptotic AdS$_5$ boundary is at $u\rightarrow\infty$.
The Hawking temperature is
\be\label{def_T}
T= {u_0\over\pi L^2}\,,
\ee
and we note that a rescaling of $T$ can be absorbed by a rescaling of the $u$ coordinate.
We work in the supergravity limit, so in particular at very large $N$, and insert a probe 
D5-brane with an $\adsBH{4}\times \sphere^2$-embedding.\footnote{D3/D5 systems at finite 
temperature and at finite chemical potential have been widely employed in applied 
holography \cite{Evans:2008nf, Filev:2009ai, Jensen:2010ga, Evans:2010hi, Grignani:2012jh}. 
For reviews see \cite{CasalderreySolana:2011us, Erdmenger:2007cm}. }
In the static gauge the D5-brane 
world-volume coordinates are $(t,x,y,u)\in \adsBH{4}\subset \adsBH{5}$ and $(\theta\,, \phi) \in \sphere^2\subset \sphere^5$, as indicated in Table \ref{table:Dbrane-construction}. 
The D5-brane profile is described by two functions, $x_\perp(u)$ and $\psi(u)$, that,
due to translation and rotation symmetries in the world-volume directions,
only depend on the radial coordinate $u$. We set $x_\perp=0$ throughout and focus 
on the angle $\psi(u)$, which controls the size of the 2-sphere wrapped by the probe 
D5-brane.\footnote{In Section \ref{sec:finite_density} we consider a D5-brane carrying
non-vanishing charge density. The ansatz $x_\perp=0$ remains consistent in this case
as well, as long as the charge density is uniform. The stability of the configuration was 
studied in \cite{Karch:2001cw}.}
The D5-brane introduces matter fields in the fundamental representation  of $\grSU(N)$, 
charged under a global $\grU(1)_B$  (baryon number), and localized in (2+1)-dimensions. 
This is the field theory we have in mind throughout the paper. 
\begin{table}[htp]
\caption{Background D-brane construction. }
\begin{center}
\begin{tabular}{c|cccccccccc}
 & $t$ & $x$ & $y$ &  $x_\perp$ & $u$& $\psi$ & $\theta$ & $\phi$ &$\tilde \theta$ &$\tilde \phi$
 \\ \hline 
$N$ D3-branes (background) & $\times$ & $\times$ &$\times$& $\times$\\ 
\hline
D5-brane (probe) & $\times$ & $\times$ &$\times$& & $\times$& &$\times$ &$\times$ \\ 
\end{tabular}
\end{center}
\label{table:Dbrane-construction}
\end{table}

For numerical computations, we find it convenient to introduce a dimensionless radial 
coordinate $\upsilon$ as follows~\cite{Babington:2003vm}:
\begin{equation}\label{upsilondef}
(u_0\upsilon)^2=u^2+\sqrt{u^4-u_0^4}\,.
\end{equation}
The horizon is at $\upsilon=1$ and the background metric (\ref{bgmetricu}) becomes
\begin{equation}\label{bgmetricrho}
ds^2=\frac{1}{2}\left(\frac{u_0\upsilon}{L}\right)^2
\left[-\frac{f^2}{\tilde{f}}\,dt^2+\tilde{f}\,(dx^2{+}dy^2{+}dx_{\perp}^2)\right]
+\frac{L^2}{\upsilon^2}\left(d\upsilon^2+\upsilon^2\,d\Omega_5^2\right)\,,
\end{equation}
where $f(\up)=1-1/\upsilon^4$ and $\tilde{f}(\up)=1+1/\upsilon^4$.
The $\ads$ boundary is at $\up\to \infty$. 

Writing $\chi(\upsilon)\equiv\cos\psi(\upsilon)$, the induced metric on the D5-brane is given by
\begin{equation}\label{gamma_5_norho}
ds^2=\frac{1}{2}\Big(\frac{u_0\upsilon}{L}\Big)^2
\bigg[{-}\frac{f^2}{\tilde{f}}\,dt^2{+}\tilde{f}\,(dx^2+dy^2)\bigg]
+\frac{L^2}{\upsilon^2}\bigg(\frac{1-\chi^2+\upsilon^2\dot{\chi}^2}{1-\chi^2}\bigg)d\upsilon^2
+L^2\Big(1-\chi^2\Big)d\Omega_2^2\,,
\end{equation}
where the dot denotes a derivative with respect to $\upsilon$.
The Euclidean DBI action for the probe D5-brane (see Appendix \ref{app:convention}) reduces to
\begin{equation}\label{SD5_gappedcharges}
I_{D5}=\mathcal{K}\,T^2 \int \dif{\upsilon}\,\upsilon^2 f 
\sqrt{\tilde{f}\left(1-\chi^2\right)\left(1-\chi^2+\upsilon^2 \dot{\chi}^{2}\right)}\,,
\end{equation}	
with the constant $\mathcal{K}$ given in \eqref{kappadef}.

The field equation for $\chi$,
\be
\ddot\chi+\frac{\upsilon  (3 \upsilon ^8+2 \upsilon ^4+3)
   \dot{\chi}^3}{(\upsilon ^8-1) (1-\chi^2)}
+   \frac{3 \chi \dot{\chi}^2}{1-\chi^2}
+\frac{2 (2 \upsilon ^8+\upsilon ^4+1) \dot{\chi}}{\upsilon  (\upsilon ^8-1)}
+\frac{2 \chi}{\upsilon ^2}=0 \,,
\label{chiequation}
\ee
can be solved numerically using standard methods. 
There are two classes of solutions with a non-trivial $\chi(\upsilon)$ profile depending on 
whether the probe D5-brane extends all the way to the horizon at $\upsilon=1$ or caps off 
outside the horizon. The former are referred to as 
``Black Hole Embedding" (BHE) solutions and the latter are so-called
``Minkowski Embedding" (ME) solutions \cite{Mateos:2006nu, Mateos:2007vn}.
A one-parameter family of black hole embedding solutions, with $0\le\chi_0\le 1$, 
is obtained by numerically integrating the field equation \eqref{chiequation} from $\upsilon=1$ 
outwards using the initial values $\chi(1)=\chi_0$, $\dot{\chi}(1)=0$. 
The condition on $\dot{\chi}(1)$ comes from requiring the field equation to be non-singular 
at the horizon. 

For the Minkowski embedding solutions, the numerical evaluation is streamlined
by a further change of variables. 
By viewing $\upsilon$ and $\psi$ as polar coordinates, the metric on $\sphere^5$ may
be rewritten as follows,
\begin{eqnarray}
d\upsilon^2+\upsilon^2\,d\Omega_5^2&=&d\upsilon^2+\upsilon^2(d\psi^2
+\sin^2\psi\,d\Omega_2^2+\cos^2\psi\,d\tilde{\Omega}_2^2)\nonumber\\
&=&dr^2+r^2\,d\Omega_2^2+dR^2+R^2\,d\tilde{\Omega}_2^2\,,
\end{eqnarray}
with
\begin{equation}\label{coord_transf_2}
\upsilon^2=r^2+R^2,\qquad r=\upsilon \sin\psi,\qquad R=\upsilon \cos\psi\,.
\end{equation}
In the new coordinates the D5-brane profile is described by a function $R(r)$ and a 
Minkowski embedding solution caps off at $r=0$.
The field equation \eqref{chiequation} becomes
\be
R''+2(R'^2+1)\left(\frac{R'}{r}
+\frac{(r R'-R)((r^2+R^2)^2+3)}{(r^2+R^2)((r^2+R^2)^4-1)}\right) =0\,,
\label{Requation}
\ee
where prime denotes a derivative with respect to $r$. 
We obtain a one-parameter
family of Minkowski embedding solutions by integrating \eqref{Requation} using the
initial values $R(0)= R_0 >1$ and $R'(0)=0$. 
The initial condition on $R'$ comes 
from requiring the field equation to be non-singular at $r=0$.

\begin{figure}
\begin{center}
\includegraphics[scale=.7]{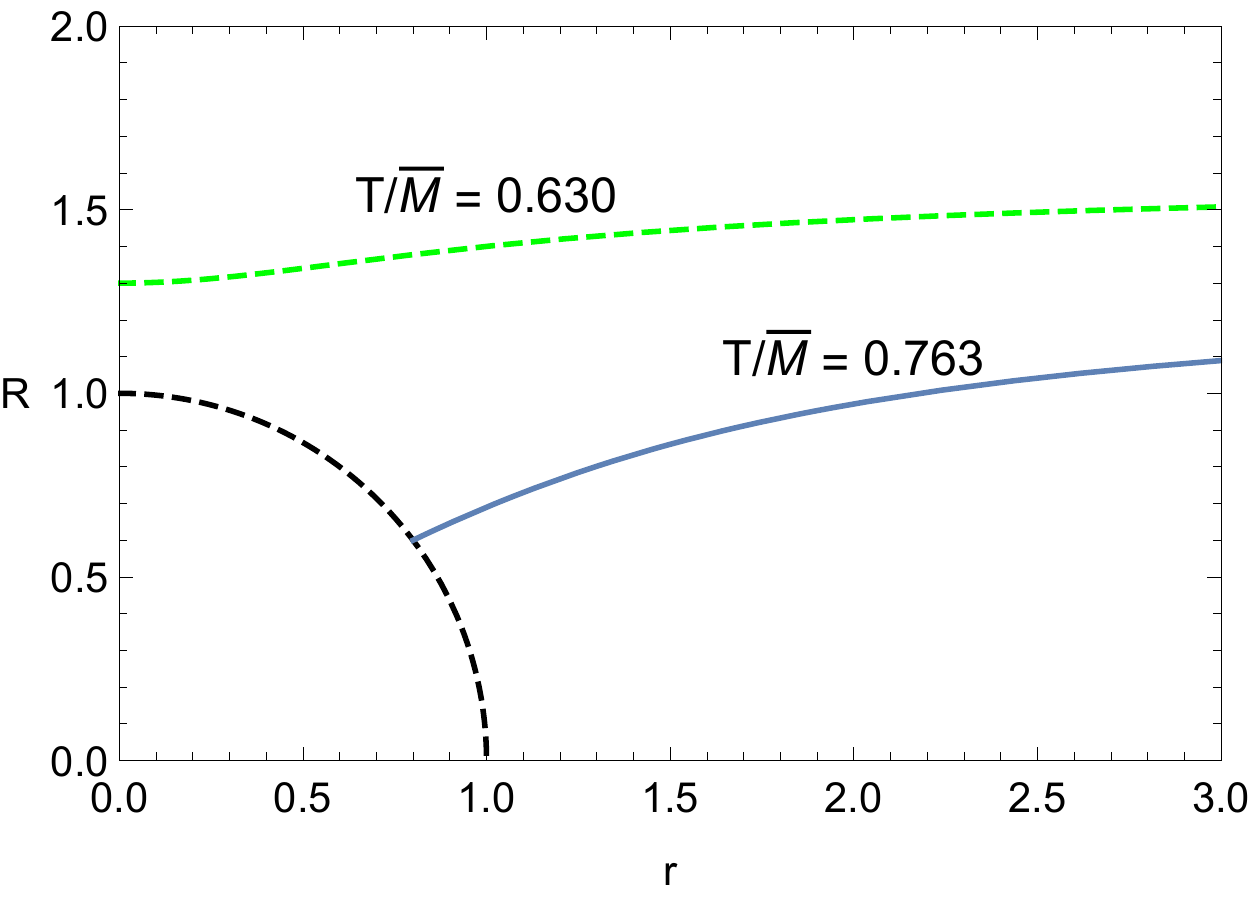}
\end{center}
\caption{ The embedding angle for the ``Minkowski embedding''  solution (dashed green curve) and the ``black-hole embedding'' solution (solid blue curve).
The dashed black curve is the horizon at $\up = 1$.
	}\label{D5_embedding}
\end{figure}

Figure \ref{D5_embedding} shows two D5-brane profiles. One is a Minkowski embedding
that ends at $r=0$ with $R_0>1$, while the other is a black hole embedding that extends 
to the horizon at $\upsilon=1$. From the figure it is clear that there exists a borderline solution 
that belongs to both embedding classes. It can either be viewed as a black hole embedding 
solution with $\chi_0=1$ that enters the horizon at a vanishing angle, or equivalently as a 
Minkowski embedding solution with $R_0=1$ that caps off at the horizon.

Each D5-brane solution is characterised by two constants, $m$ and $c$, that can be read
off from the asymptotic behavior at the boundary,   
\be
\label{psi_boundary}
 \chi(\up) \sim {m\over \up}+{c\over \up^2}+\dots \,, &&\quad \up\to \infty\,, \qquad \text {BHE}\,,\\ \nn
R (r)\sim m + {c\over r} +\dots \,,\>\>\> && \quad r\to \infty\,,  \qquad \text{ME}\,.
\ee 
They represent the boundary mass $M_b$ and the condensate 
$\langle \mathcal O_b \rangle$ of $\grU(1)_B$ flavour charged degrees of freedom in the 
dual field theory~\cite{Mateos:2006nu, Mateos:2007vn, Kobayashi:2006sb},
\be\label{rel_m_T}
M_b = {u_0\over 2\sqrt 2 \pi \ell_s^2} m \,, 
\qquad 
\langle \mathcal O_b \rangle =- {(4\pi \ell_s)^2\over \sqrt 2}\, T_5 \, u_0^2\, c\,. 
\ee
Note that our definition of the boundary mass $M_b$ differs from that in \cite{Iqbal:2014cga} by a factor of $\frac{\sqrt{\lambda}}{2 \pi}$.
By using \eqref{def_T} and \eqref{braneparameters}, we see that the mass parameter $m$ read 
off from our numerical solutions is proportional to the 
scale invariant ratio of the boundary mass and the temperature,
\be
\label{invtemp}
m = {2 \sqrt 2 \over \sqrt \lambda} \, {M_b \over T}\equiv {\bar M\over T} \,.
\ee
In the following, we will use $m$ as a measure of the (inverse) temperature at fixed $\bar M$.
Note that the trivial constant profile, $\chi(\upsilon)=0$ ($\psi={\pi/2}$), is a solution of the field equation \eqref{chiequation} at any temperature and corresponds to $m=0$.

\begin{figure}
\begin{center}
\includegraphics[scale=.82]{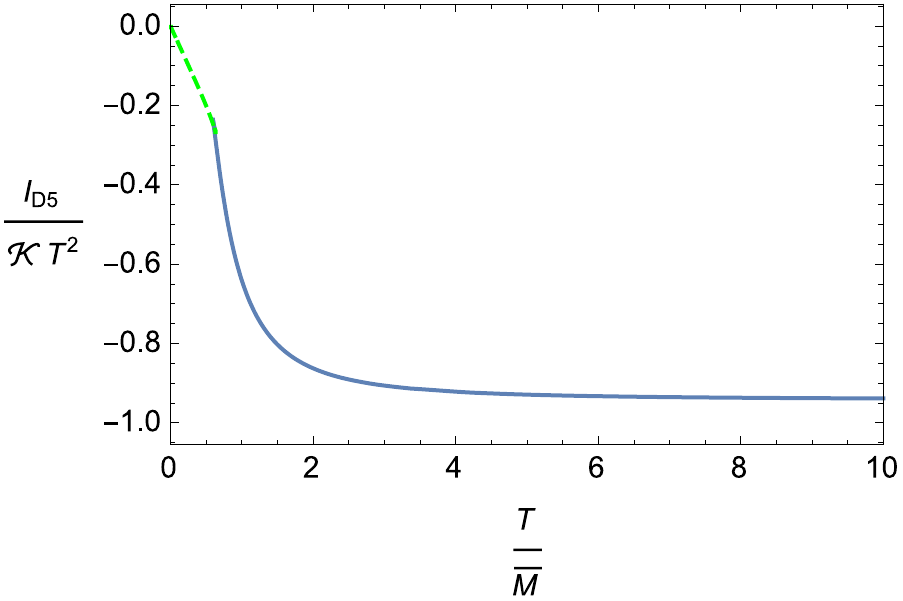}\quad
\includegraphics[scale=.82]{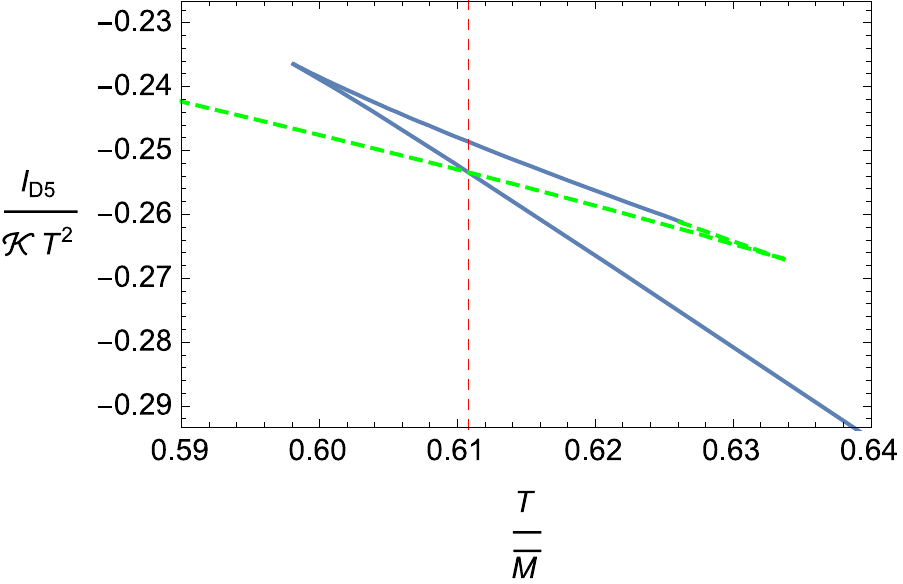}
\end{center}
\caption{Regularised free energy for a D5-brane in D3-branes background. The dashed green 
(solid blue) curves correspond to Minkowski (black hole) embeddings. The figure on 
the right shows an expanded view of the region near the phase transition. The dotted 
vertical red line indicates the critical temperature $T_c/\bar{M}=0.611$.}
\label{D5_action}
\end{figure}

In order to determine the thermodynamically stable D5-brane solution at a given temperature, 
we compare the regularised free energy of the different solutions.  
We review the main points 
of the regularisation procedure worked out in \cite{Karch:2005ms} in 
Appendix~\ref{app:thermodynamics-D5} and the resulting free energies are shown in 
Figure~\ref{D5_action}.  
The low-temperature phase, {\it i.e.} low $T/\bar M$, corresponds to a 
Minkowski embedding (or ``gapped'') solution, where the two-sphere shrinks down at a finite 
distance away from the horizon ($R(0)>1$) and the spectrum of ``quark-antiquark'' bound states has 
a mass gap~\cite{Mateos:2006nu, Mateos:2007vn, Albash:2006ew}.
At high $T/\bar M$ the D5-tension can no longer balance the gravitational attraction of the 
background D3-branes and the favored solution is a black hole embedding solution, 
dual to a gapless meson spectrum in the boundary field theory. 
There is a phase transition between the two types of embeddings. The 
right-hand plot in Figure~\ref{D5_action} zooms in on the region near the critical 
temperature and reveals the characteristic swallow tail of a first-order transition. This is a 
universal feature of all D$p$-D$q$ systems~\cite{Mateos:2006nu, Mateos:2007vn}.

The free energy shown in Figure~\ref{D5_action} is obtained from numerical 
solutions of the field equations \eqref{chiequation} and \eqref{Requation} at different 
temperatures. In Appendix~\ref{asymptotia} we consider its high and low temperature limits, 
where the field equations simplify and analytic results can be obtained. 

\subsection{Monopole two-point function}
\label{sec:monopole-nocharge}

In order to calculate the two-point correlation function of monopole operators in the
dual field theory we consider a probe D3-brane on top of the background 
D3-D5-brane system. The probe D3-brane ends on the D5-brane and 
thus appears as a magnetically charged object in the D5 
worldvolume \cite{Strominger:1995ac}. Furthermore, 
the D3-brane is embedded in $\adsBH{5}\times \sphere^5$ in such a way that it wraps 
the same $\sphere^2\subset \sphere^5$ as the D5-brane does and extends along
a one-dimensional curve in $\adsBH{4}\subset\adsBH{5}$ with endpoints at the $\ads_4$ 
boundary. This accounts for three out of four 
of the D3 worldvolume directions. The remaining worldvolume direction is transverse to 
the D5-brane, along $\psi$ in the parametrisation \eqref{bgmetricu}, from $\psi=0$ 
to $\psi(u)$ where the D3-brane ends on the D5-brane. The probe D3-brane thus
appears as a particle in $\adsBH{4}$ and is magnetically charged under 
the D5 world-volume gauge field $A$, {\it i.e.}\ it models a bulk magnetic 
monopole \cite{Iqbal:2014cga}. 
 
The D3-brane fills part of the $\sphere^3\subset \sphere^5$ described by 
$(\psi,\theta,\phi)$ in \eqref{bgmetricu}. The fraction of the $\sphere^3$ volume 
that is filled depends on the D5 embedding. At the boundary, the D5-brane is at
$\psi={\pi\over 2}$ and the D3-brane fills half of the $\sphere^3$. 
The curve connecting the insertion points on the $\ads_4$ 
boundary extends into the $\adsBH{4}$ bulk, where the 
D5-brane generically moves away from $\psi={\pi\over 2}$ and the D3-brane 
occupies a smaller fraction of the $\sphere^3$ volume. In particular, if the 
curve extends to where the D5-brane caps off in a Minkowski embedding then 
the volume of the D3-brane shrinks to zero at that point.  
 
Computing the two-point boundary monopole 
correlation function in the large $N$-limit amounts to evaluating the corresponding 
on-shell D3-brane action as a function of the separation between the brane endpoints 
on the $\ads_4$ boundary. 
In Section~\ref{sec:finite_density} we present results from 
a numerical evaluation of the equal time two-point 
monopole correlator at finite temperature and background charge density, generalising 
the zero-temperature results obtained in \cite{Iqbal:2014cga}. We begin, however, 
with the simpler case of vanishing charge density at finite temperature. 

We find it convenient to use D3-brane world-volume coordinates $(s,\chi, \theta, \phi)$
that match the coordinates we used for the D5-brane embedding in Section~\ref{sec:preliminaries}. 
Here $s$ parametrises the curve $\{\upsilon(s),x(s),y(s)\}$ traced out by the probe 
D3-brane in the $\adsBH{4}$ part of the background geometry 
\eqref{bgmetricrho}. This curve is spacelike when we consider a two-point function 
of monopole operators inserted at equal time on the boundary.\footnote{In this case the
probe D3-brane is strictly speaking a D3-instanton.}
The variable $\chi=\cos\psi$ is restricted to the range $\chi_{D5}\leq \chi \leq 1$, where
$\chi_{D5}=\chi\big(\upsilon(s)\big)$ corresponds to the intersection between the D5- and 
D3-brane worldvolumes.

\begin{figure}
\begin{center}
\subfigure[Bulk monopole effective mass]{\includegraphics[scale=.78]{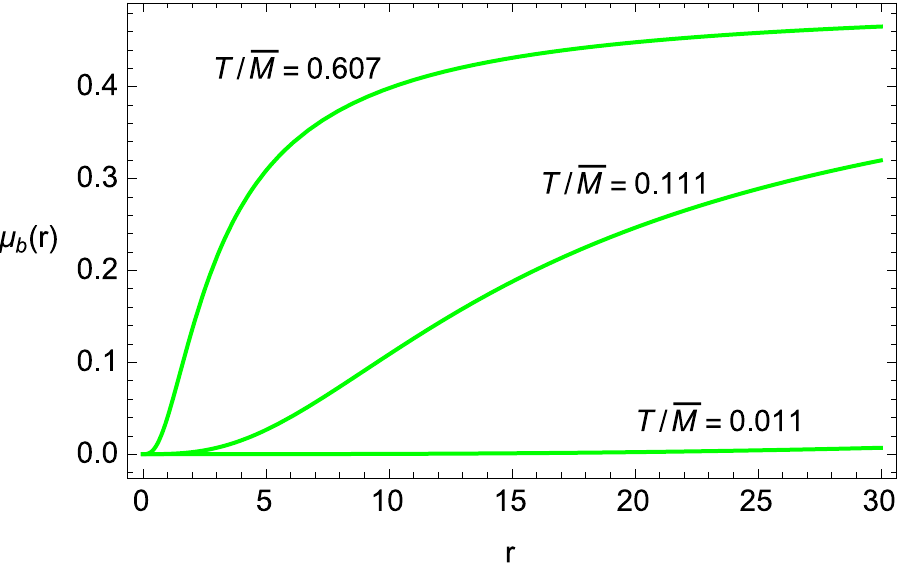} 
\quad \includegraphics[scale=.78]{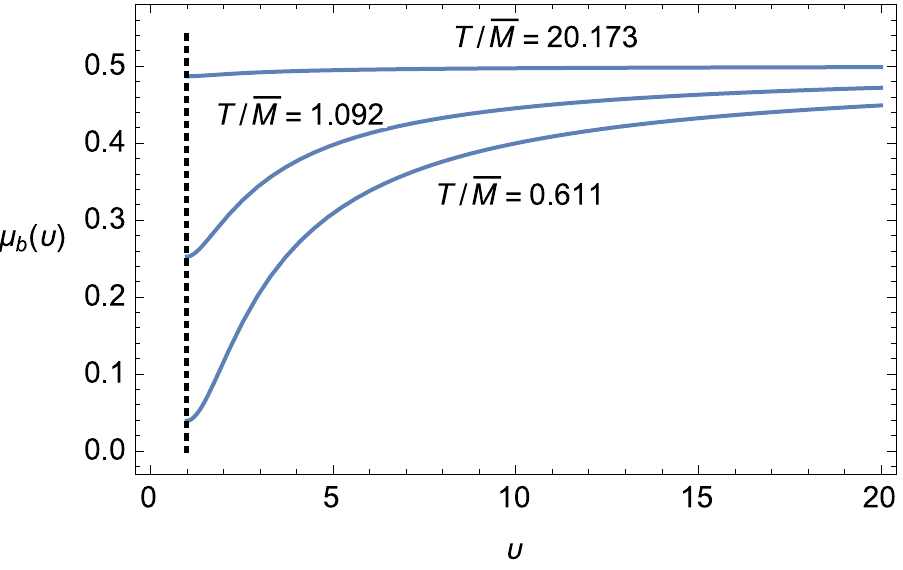}  \label{fig:mu-hMink} }\qquad
\subfigure[Regularised D3-brane action]{ \includegraphics[scale=0.78]{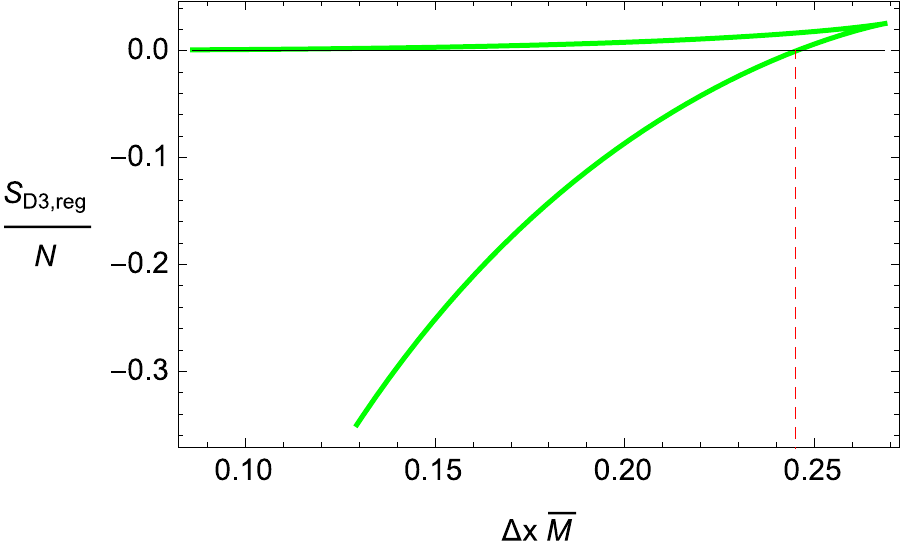}  
\quad \includegraphics[scale=0.5625]{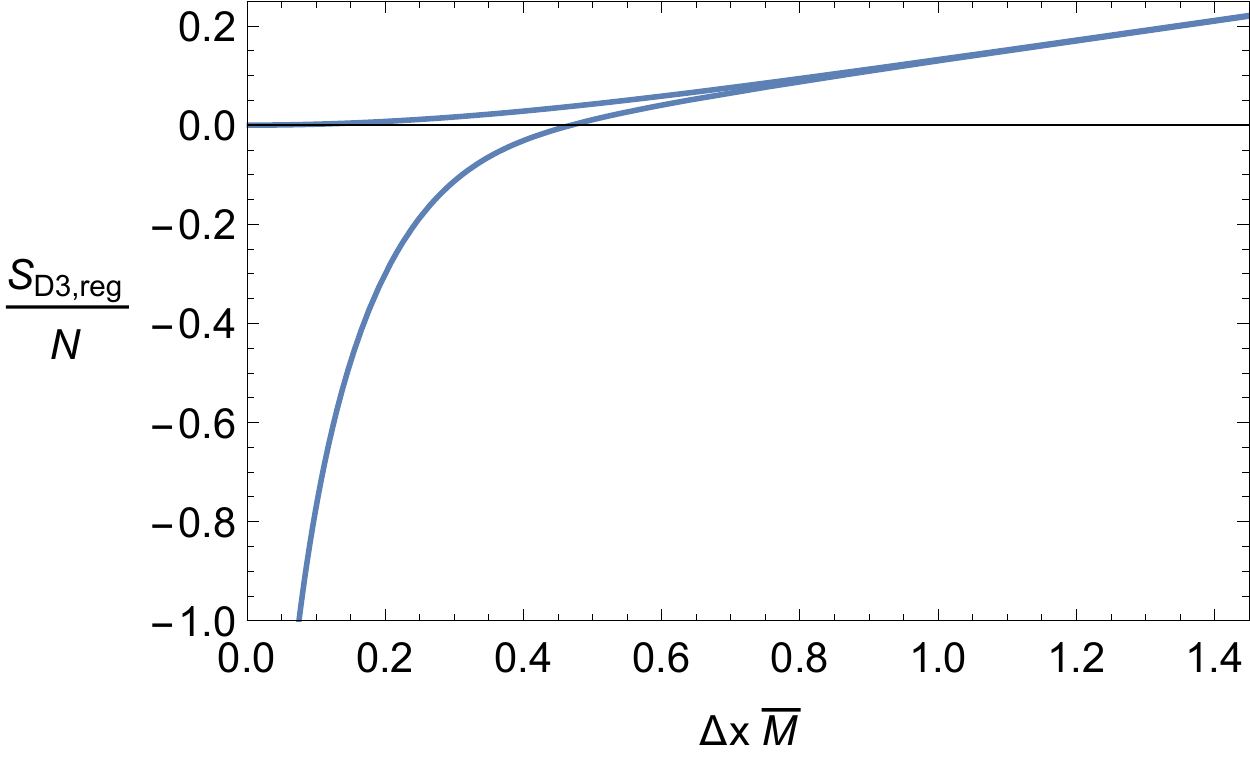} \label{fig:ungapped}} 
\end{center}
\caption{(a) Bulk monopole effective mass for Minkowski (black hole) embeddings 
on the left (right) at different temperatures; (b) Regularised D3-brane action versus 
$\Delta x$ for Minkowski (black hole) embeddings 
on the left (right) at $T/\bar{M}=0.1 \,(0.68)$. The dashed vertical line in the left panel
indicates the critical endpoint separation marking the transition from a single ``hanging"
geodesic to two disconnected vertical geodesics in a Minkowski embedding.
}
\label{fig:Sd3vsDx-norho}
\end{figure}

In the charge neutral case, the action \eqref{general_D3_action} for a probe D3-brane 
only contains the DBI term. 
Upon integrating over the coordinates
$(\chi,\theta,\phi)$ the DBI action reduces to that of a point particle,
\be\label{dbi_particle}
S_{D3} = N \int_{\CC} \dif{s} \, {m_{b}(s)} \, \sqrt{G_{xx} \left({\dot x(s)}^2
+{\dot y(s)}^2\right)+G_{\up\up} {\dot \up(s)}^2},
\ee
where $G_{IJ}$ is the pull-back of the ten-dimensional spacetime metric to the D3-brane 
world-volume, a dot indicates a derivative with respect to $s$, and 
$m_{b}(\up)$ is a position dependent mass given by
\be\label{mexp_mb}
m_{b}(\up)\equiv{\mu_b (\up)\over L} \equiv {2\over \pi L}  \int_{\chi_{D5}}^1 
\sqrt{1-\chi^2} \,\dif{\chi}\,. 
\ee
We refer to the dimensionless quantity $\mu_b (\up)$ as the effective 
mass of the bulk monopole.\footnote{Note that our normalisation convention for  
$\mu_b$  differs from that of \cite{Iqbal:2014cga} by a factor of $N$.}
It follows that the dynamics of the probe D3-brane depends on the embedding of the 
D5-brane it ends on.
In a Minkowski embedding $\mu_b(\up)$ shrinks to zero at the point where the D5 caps off,
while in a black hole embedding $\mu_b(\up)$ remains non-zero all the way to the horizon.
This is clearly visible in Figure \ref{fig:mu-hMink}, which shows $\mu_b$ as a function
of position at different temperatures.

For the actual computation, it is convenient to absorb the effective mass into the 
induced metric as a conformal factor and define a rescaled metric~\cite{Iqbal:2014cga},
\be\label{gbar_mb}
\tilde G_{IJ} = m_{b}^2(\up) \, G_{IJ}\,.
\ee
The on-shell D3-brane action is then given by the length of a geodesic in the rescaled metric
connecting the monopole operator insertion points at the $\ads_4$ boundary, which can
without loss of generality be assumed to lie on the $x$-axis.
The geodesic extends along $\{\up(s),x(s)\}$ and intersects the boundary at
$\up\to \infty$, $x\to\pm {\Delta x\over 2}$. It has a turning point at $x=0$,
$\up=\up_*$, where ${d\up\over ds}=0$. 

As shown in Appendix~\ref{sec:D3_general}, the D3-brane action \eqref{dbi_particle} 
can be re-expressed as
\be\label{Sd3-norho-mainbody}
{S_{D3}} 
=2 N\int_{\upsilon_*}^\infty \dif{\upsilon} \sqrt{{\tilde G_{\upsilon\upsilon}\over 1-P^2 \tilde G^{xx}}} 
= 2 N \int_{\upsilon_*}^\infty \dif{\upsilon} {\mu^2_b(\up) \over  \sqrt{\upsilon^2\mu^2_b(\up)
- 2 \bar P^2\tilde f^{-1}(\up)}}\,,
\ee
where $P$ is the conserved charge associated with translation invariance along the 
spatial $x$-direction, and $\bar P$ is the corresponding dimensionless variable,
\be\nn
P\equiv \dot x\, \tilde G_{xx}\,, \qquad \bar P= {P \over \pi\, T}\,. 
\ee
The separation of the D3 endpoints at the boundary is given by
\be\label{dx-norho-mainbody}
\Delta x 
= 2 \int_{\upsilon_*}^\infty  \,\sqrt{{\tilde G_{\upsilon\upsilon}\,\over 1-P^2 \tilde G^{xx}}} \,\tilde G^{xx}\, P\, \dif{\upsilon}\, ,
\ee
which can in turn be expressed in terms of dimensionless quantities as
\be\label{dx-norho-invariant}
\Delta x\, \bar M= {4m\over \pi} \int_{\upsilon_*}^\infty 
{\dif{\upsilon}\over \upsilon^2\tilde f(\up)}
{ \bar P\over \sqrt{\upsilon^2\mu^2_b(\up)- 2 \bar P^2\tilde f^{-1}(\up)}} \, .
 \ee
In our numerical computation, $\bar P$ is an input parameter and we evaluate both the
D3-brane action and the endpoint separation as a function of $\bar P$ for a given D5-brane
embedding solution. 

The location of the turning point, $\up=\up_*$ of a geodesic with 
$\bar P\neq 0$ depends on the D5-brane embedding. 
The condition for having a turning point is
\be\label{tp_location}
\tilde G_{xx}(\up_*)= P^2 \,,
\ee
or equivalently 
\be\label{turning-point-norho}
\up_* = \sqrt{{\bar P^2\over \mu^2_b(\up_*)}
+\sqrt{{\bar P^4\over \mu^4_b(\up_*)}-1}}\,. 
\ee
A real valued solution requires $\vert\bar P\vert\ge \mu_b(\up_*)$. 
In a Minkowski embedding, this condition is always satisfied for some value of $\up_*$ 
on the D5-brane because $\mu_b(\up)$ goes to zero as the D5-brane caps off. 
In addition to geodesics with turning points, the Minkowski embedding 
supports a $\bar P=0$ geodesic that extends ``vertically" from the boundary to the point 
where the D5-brane caps off, depending on the type of D5-brane embedding. A pair of 
such vertical D3-branes turns out to be the thermodynamically favoured configuration at 
sufficiently large endpoint separation when the probe D5-brane is in a Minkowski 
embedding.

In a black hole embedding, on the other hand, the geodesic may reach the
horizon at $\up=1$ before the turning point condition \eqref{turning-point-norho} is satisfied.
In this case, the geodesic instead turns around at the horizon. This is not immediately 
apparent in the $\up$~coordinate, because the coordinate transformation \eqref{upsilondef}
is degenerate at the horizon, but by going back to the original 
$u$~coordinate it is straightforward to show that the geodesic is quadratic in $x$ near the
horizon,
\begin{equation}\label{new_branch}
\frac{u(x)}{u_0} = 1 + \Big(\frac{\mubh^2}{\bar P^2}-1\Big)\frac{u_0^2 x^2}{L^4} + \order(x^4),
\end{equation}
when $\vert\bar P\vert < \mubh$, where $\mubh$ is $\mu_b(\upsilon)$ evaluated at the horizon $\upsilon = 1$.
This behaviour is also apparent in numerical solutions of the geodesic equation at sufficiently 
low $\bar P$. 

The D3-brane action $S_{D3}$ is in fact divergent for all 
the geodesic curves we have described. The divergence comes from the region near 
the boundary $\up\to \infty$. We regularise it by introducing an upper cut-off at 
$\up=\up_{max}\gg 1$ in \eqref{Sd3-norho-mainbody} and
subtracting the action of a geodesic in the $\bar P \rightarrow 0$ limit,
\be\label{sd3-disc-norho}
{S_{D3}^{\,0}}=2N \int^{\up_{max}}_{\up_{min}} \dif{\up} \sqrt{\tilde G_{\up\up}}\, 
= 2N \int^{\up_{max}}_{\up_{min}} \dif{\up}\,{\mu_b(\up)\over \up}\, ,
\ee
where $\up_{min}$ depends on the D5-brane embedding. 
For a black hole embedding it is at the horizon, 
$\up_{min}=1$, while for a Minkowski embedding it is where the D5-brane caps off.
In this case, we can use the coordinates $\{ r\,, R\}$ introduced in \eqref{coord_transf_2} 
and set the lower limit of the radial variable in the integral to $r_{min}=0$. 
By this convention, a disconnected configuration in Minkowski embedding, where two 
separate vertical D3-branes extend from the boundary, has vanishing regularised action. 

In Figure~\ref{fig:Sd3vsDx-norho}(b) we plot the regularised action, 
\be
\label{eq:SD3reg}
S_{D3, \rm reg} \equiv S_{D3}-S_{D3}^{\,0}\,,
\ee
obtained by numerically evaluating the integrals in \eqref{Sd3-norho-mainbody} and 
\eqref{sd3-disc-norho} for different values of the dimensionless parameter $\bar P$,
against the endpoint separation $\Delta x$ obtained at the same value of $\bar P$. 
For a Minkowski embedding the system undergoes a first order phase 
transition, similar to the zero-temperature case studied in \cite{Iqbal:2014cga},
at a critical value of the endpoint separation, indicated by a dashed vertical line 
in the figure.
At small values of the endpoint separation $\Delta x$, the thermodynamically
stable branch consists of connected solutions with large $\vert\bar{P}\vert$, 
for which the turning point is located far from the cap of the D5-brane.  
Following this branch towards smaller $\vert\bar{P}\vert$, the turning point moves 
deeper into the bulk geometry while both the endpoint separation
$\Delta x$ and the regularised free energy \eqref{eq:SD3reg} increase. Eventually
the free energy becomes positive and this branch is disfavoured compared to a 
disconnected branch with two separate vertical D3-branes. 
The critical endpoint separation is indicated by a dashed vertical line in the figure. 

Following the (now unstable) connected branch to smaller $\vert\bar{P}\vert$, the 
endpoint separation $\Delta x$ increases towards a local maximum, which marks
the endpoint of this branch of solutions. Even smaller values of $\vert\bar{P}\vert$ 
give rise to a third branch of solutions where the endpoint separation decreases 
from its local maximum and eventually approaches zero in the limit of vanishing 
$\vert\bar{P}\vert$. On this branch the turning point continues to move deeper into 
the bulk as $\vert\bar{P}\vert$ is decreased and touches the cap of the D5-brane
precisely when $\vert\bar{P}\vert=0$.
The small $\vert\bar{P}\vert$ branch of solutions is always disfavoured compared 
to the other two branches. This is apparent from our numerical results but can also 
be seen by expanding the integrands in \eqref{Sd3-norho-mainbody} and 
\eqref{dx-norho-invariant} in powers of $\bar P$ for very low $\bar P$. The endpoint 
separation is a linear function of $\bar P$, while the regularised action is a quadratic 
function of $\bar P$, and thus of $\Delta x$. It follows that the regularised action is 
always positive on this branch.\footnote{Such a low momentum expansion is explicitly
carried out for the more general case with non-zero background charge density 
in Appendix \ref{app:low-P-limit}.}
\begin{figure}
\begin{center}
\includegraphics[scale=.78]{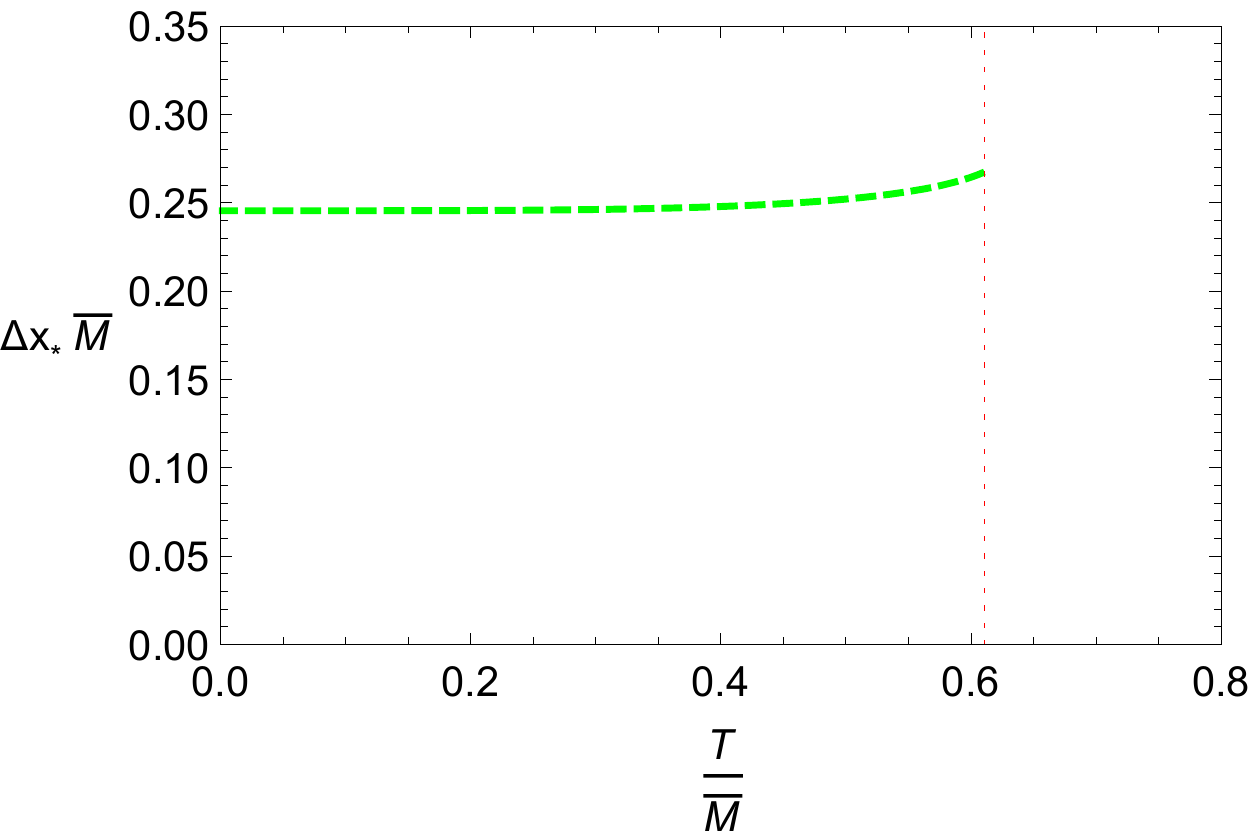}
\end{center}
\caption{Temperature dependence of $\Delta x_{*}$, the critical separation between 
D3-brane endpoints in a Minkowski embedding, beyond which the thermodynamically 
favoured configuration consists of two separate segments extending vertically to 
where the D5-brane caps off.
Above the critical temperature $T_c/\bar M=0.611$ (indicated by the dotted vertical
line) the D5 brane is in a black hole embedding where there is no disconnected 
D3-brane configuration.
}\label{DeltaXvsT}
\end{figure}

For a black hole embedding the story is rather different. In this case the turning point 
of a ``hanging" geodesic can be either outside the black hole or exactly at the horizon 
$\up=1$. The two possibilities give rise to two branches of D3-brane solutions and 
the branch that is thermodynamically stable at all values of the endpoint separation 
$\Delta x$ turns out to be the one where the turning point is outside the 
horizon.\footnote{In a black hole embedding there is no analog of the disconnected 
branch, with separate D3-brane segments extending from the boundary to where the 
D5-brane caps off, that dominates at large endpoint separation in a Minkowski 
embedding. If a D3-brane were to end at the horizon its boundary would include an 
$\sphere^2$ of finite area that is not inside the D5-brane world-volume. 
We thank N.\ Iqbal for pointing out an error in a previous version of the paper
concerning this issue.}

The right panel in Figure~\ref{fig:Sd3vsDx-norho}(b) shows our numerical results for 
the regularised D3-brane action as a function of the endpoint separation in a black 
hole embedding.
Following the stable branch towards larger $\Delta x$ one finds 
a critical value of the conserved charge $\vert\bar{P}\vert=\mubh$ for which the
geodesic touches the horizon at the turning point. At the critical value of $\vert\bar{P}\vert$
the $O(x^2)$ term in \eqref{new_branch} vanishes. The unstable branch corresponds 
to $\vert\bar{P}\vert$ below the critical value
and geodesics that turn around at the horizon. 

Both the stable and unstable branches extend to infinite endpoint separation. 
To see this, consider the integrals in \eqref{Sd3-norho-mainbody} and 
\eqref{dx-norho-invariant} precisely at the critical value 
$\vert\bar{P}\vert=\mubh$. It is straightforward to establish that both integrals 
diverge logarithmically in this case, with the divergence coming from the low end of
the $\up$ integration, near the horizon. The divergence can, for instance, be regulated
by introducing a cutoff at $\up=1+\epsilon$ and in the limit of $\epsilon\ll 1$ one
easily finds that\footnote{A detailed analysis of these divergences, including the 
more general case at non-vanishing charge density, is presented in 
Appendix~\ref{sec:highTmonopole}. }
\be
\frac{S_{{D3,\rm reg}}}{N} \approx \pi T \mubh \Delta x 
\xrightarrow{T \rightarrow \infty} \frac{\pi}{2} T \Delta x.
\label{eq:largeDeltax}
\ee 
The monopole two-point function \eqref{monopole-corr} at high temperature
is thus exponentially suppressed at large spatial separation with a characteristic
length scale that scales inversely with temperature. This is the expected 
behaviour of a thermally screened system. At low temperatures, on the other 
hand, the D5-brane is in a Minkowski embedding and the favoured
D3-brane configuration at large endpoint separation is a pair of disconnected
vertical segments, for which the monopole two-point function is a constant
independent of $\Delta x$. This signals the condensation of monopoles at low 
temperatures in this system. The critical endpoint separation, at which the 
disconnected configuration becomes dominant in the low-temperature 
Minkowski embedding phase, has a weak temperature 
dependence shown in Figure~\ref{DeltaXvsT}.

\section{Finite charge density phase}
\label{sec:finite_density}

\subsection{Thermodynamics of charged D5-branes}
\label{d5thermo}

By turning on a $\grU(1)$ gauge field on the D5-brane we can generalise the results
of the previous section to study monopole correlation functions in a compressible Fermi-liquid 
phase at finite charge density. We begin by giving a quick overview of the resulting charged 
D5-brane thermodynamics before turning our attention to the monopole correlators. 
Our discussion of D5-brane thermodynamics parallels that of \cite{Kobayashi:2006sb}, 
which considered charged D7-branes in a D3-brane background.

The Euclidean action for D5-brane at finite charge density is worked out in 
Appendix~\ref{app:convention}. For convenience, we repeat the final expression
\eqref{d5charged} here,
\be\label{routhian}
I_{D5}=
 \mathcal{K} \,T^2 \int \dif{\up}\, \up^2\, f\,
 \sqrt{\tilde f(1-\chi^2) \left(1-\chi^2+\up ^2 \dot{\chi}^2\right)} 
 \sqrt{\frac{Q^2}{\tilde f^2 \up ^4 \left(1{-}\chi^2\right)^2}+1}\,.
\ee
The induced metric on the D5-brane is still parametrised as in \eqref{gamma_5_norho} 
and the finite charge density enters via the dimensionless parameter $Q$ in the action 
(see Appendix~\ref{app:convention} for details).
The field equation for $\chi$, obtained by varying \eqref{routhian}, can be solved 
numerically using the same methods as employed for the charge-neutral case in
Section~\ref{sec:preliminaries}. 
In the absence of explicit bulk sources, electric field lines emanating from the $\ads_4$ boundary 
have nowhere to end if the D5-brane caps off before the horizon \cite{Kobayashi:2006sb}. 
The Minkowski embedding solutions are therefore unphysical at finite charge density and
the only consistent solutions are black hole embeddings. 

\begin{figure}
\begin{center}
\subfigure[Low temperature ]{\includegraphics[scale=.78]{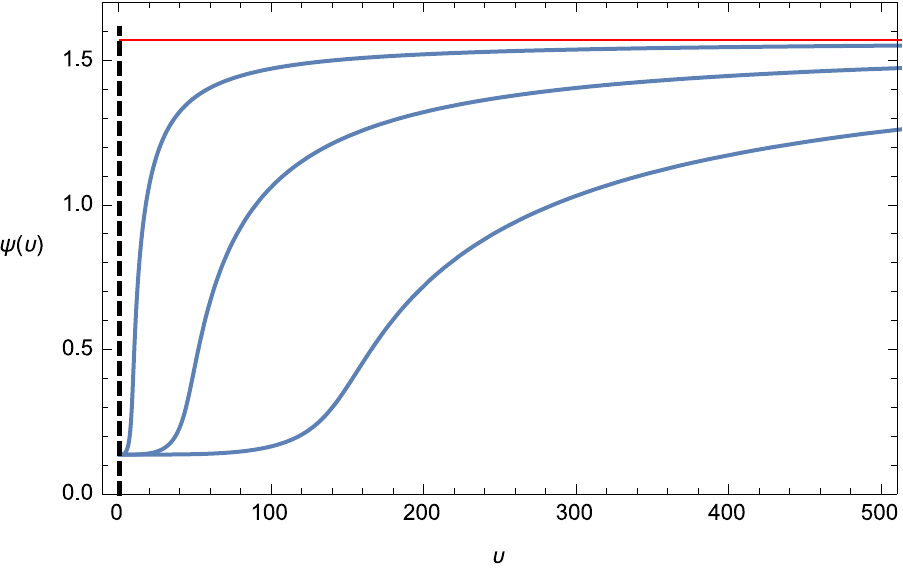}}
\quad\subfigure[High temperature ]{\includegraphics[scale=.78]{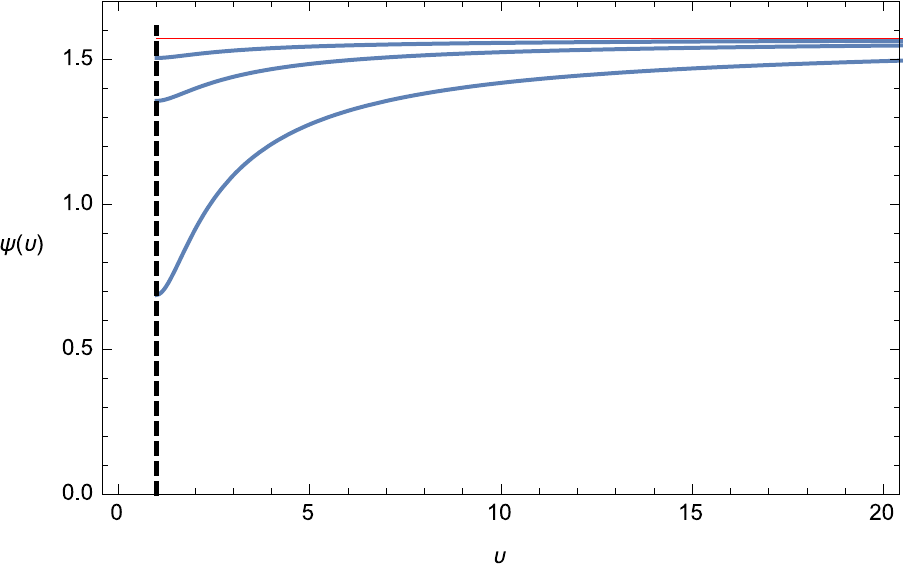}}
\end{center}
\caption{D5-brane solutions at constant charge density $\bho=0.02$ for different temperatures. 
The red line represents constant embedding while the blue lines describe the BH embeddings for a) $T/\bar{M}=0.1,\,0.02,\,0.006$ and b) $T/\bar{M}=6.323,\,2,\,0.577$ from top to bottom.
The dashed black line indicates the horizon at $\upsilon=1$.}\label{fig:psiVsrhoQ01}
\end{figure} 

The relevant variables when it comes to the physical interpretation and presentation 
of our results are the temperature $T$ and the charge density $\rho$ in the boundary 
theory.\footnote{The physical charge density $\rho$ is related to the temperature and 
the dimensionless parameter $Q$ appearing in the action through \eqref{def_rho}.}
We fix the overall scale by working at a fixed boundary mass $\bar M$ and express our
results in terms of dimensionless combinations,
\begin{equation}
m= \frac{\bar M}{T} \,,\qquad \qquad \bho = \frac{\rho}{N \bar{M}^2}\,.
\end{equation}
The phase diagram of the model is mapped out by separately varying $\bho$ and $m$.
In particular, if we keep $\bho$ fixed and consider very high temperature we 
expect thermal effects to swamp any effect of the charge density while in the limit
of low temperature the finite charge density should dominate. 
This is readily apparent in our numerical results, but we also demonstrate it explicitly 
by considering the different asymptotic limits of parameters in Appendix~\ref{asymptotia}.

Numerical solutions for $\chi$ are obtained by integrating the field equation 
outwards from the horizon, with the charge density $Q$ and
the boundary value at the horizon $\chi_0=\chi(1)$ as dimensionless input parameters.
For given values of the input parameters in the range $0 \leq \chi_0 \leq1$ and $Q \geq 0$, 
the inverse temperature $m(\chi_0,Q)$ can be read off from the asymptotic behaviour of the 
numerical solution as in \eqref{psi_boundary}. 
The charge density $\bho(\chi_0,Q)$ is then easily determined 
using the relation $\bho\,m^2=Q/2$. This procedure uniquely determines the physical
variables $m$ and $\bho$ as functions of the numerical input parameters $\chi_0$ and $Q$.
The inverse mapping $(m,\bho)\rightarrow (\chi_0,Q)$ is not single valued, however, and 
this leads to phase transitions as was already seen in the zero-charge case in 
Section~\ref{sec:preliminaries}. The constant $\chi=0$ solution is also present and 
can be viewed as the high-temperature limit of a black hole embedding, as is
apparent in Figure~\ref{fig:psiVsrhoQ01}.

\begin{figure}
\begin{center}
\subfigure[]{\includegraphics[scale=.55]{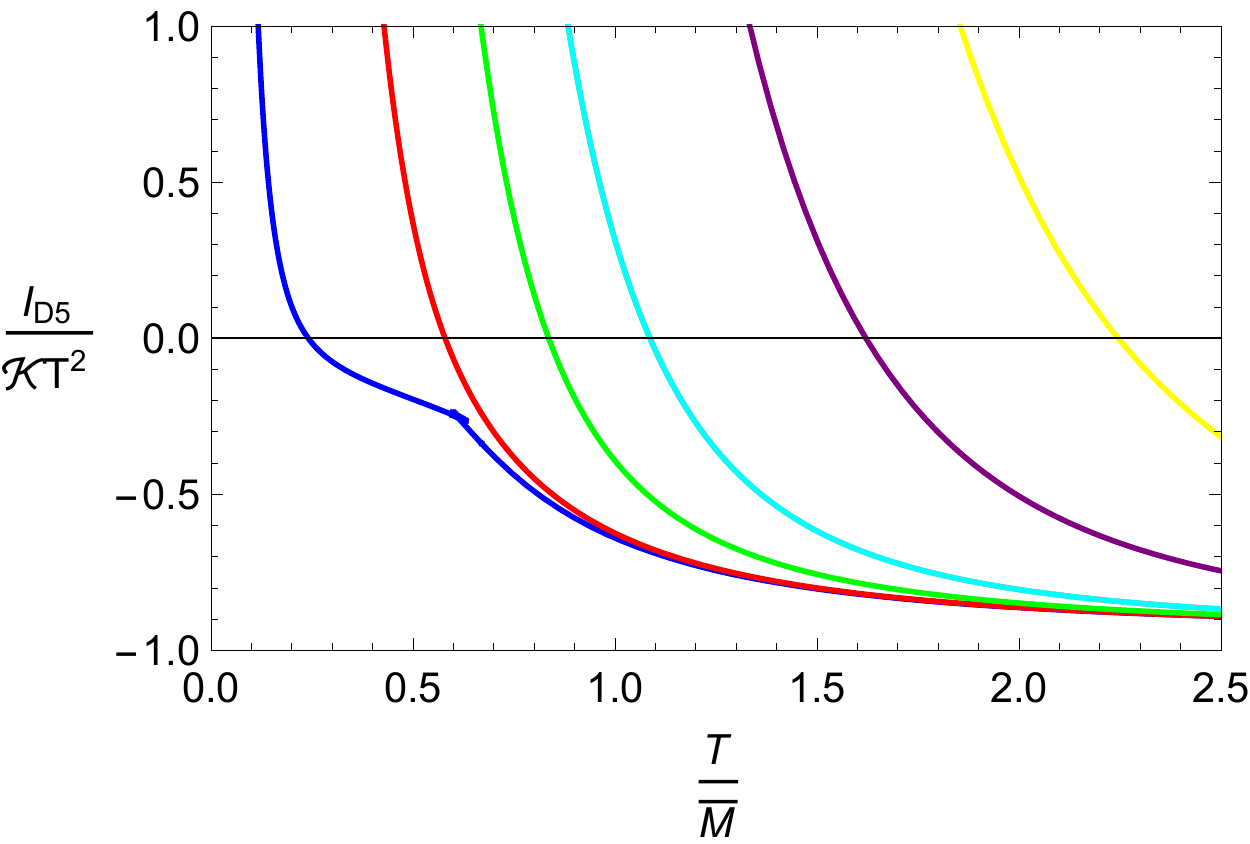}\quad}
\subfigure[]{\includegraphics[scale=.5]{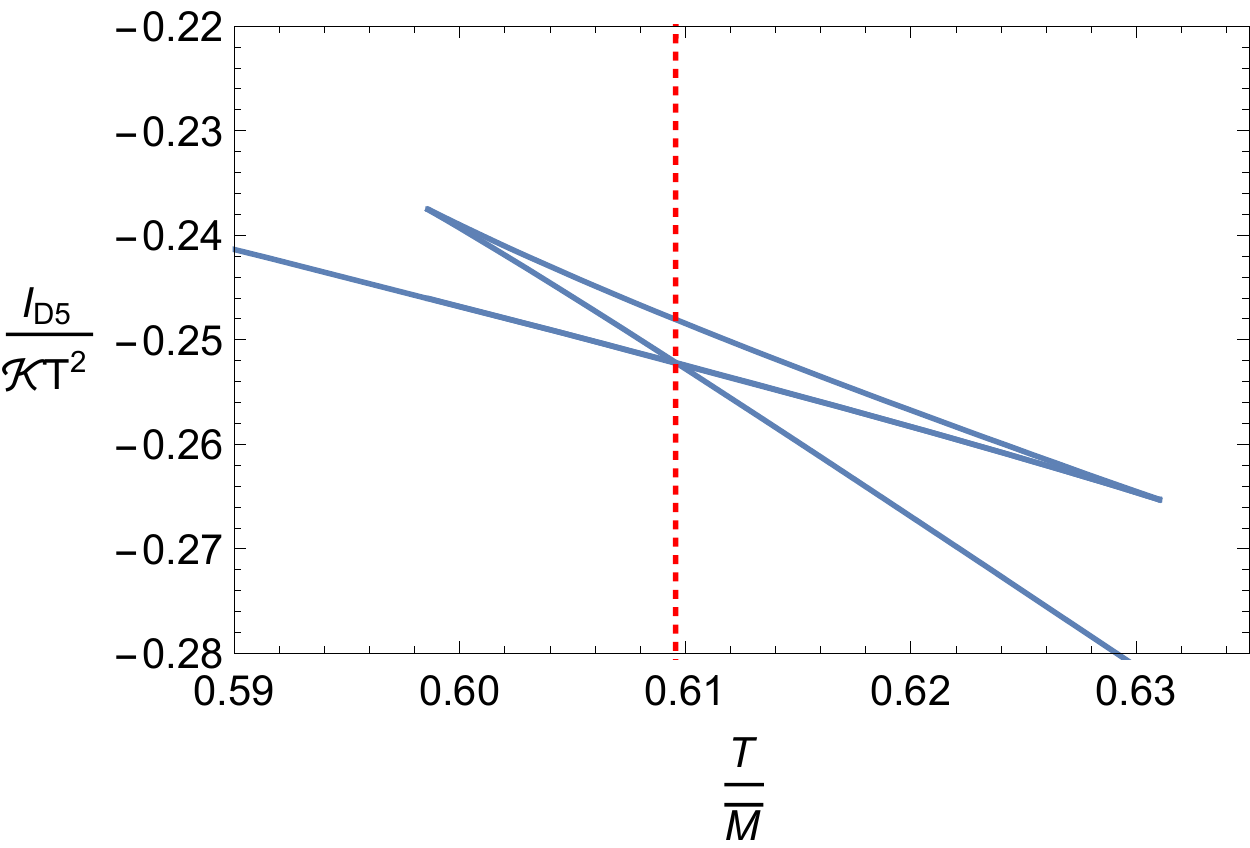}\quad
\includegraphics[scale=.5]{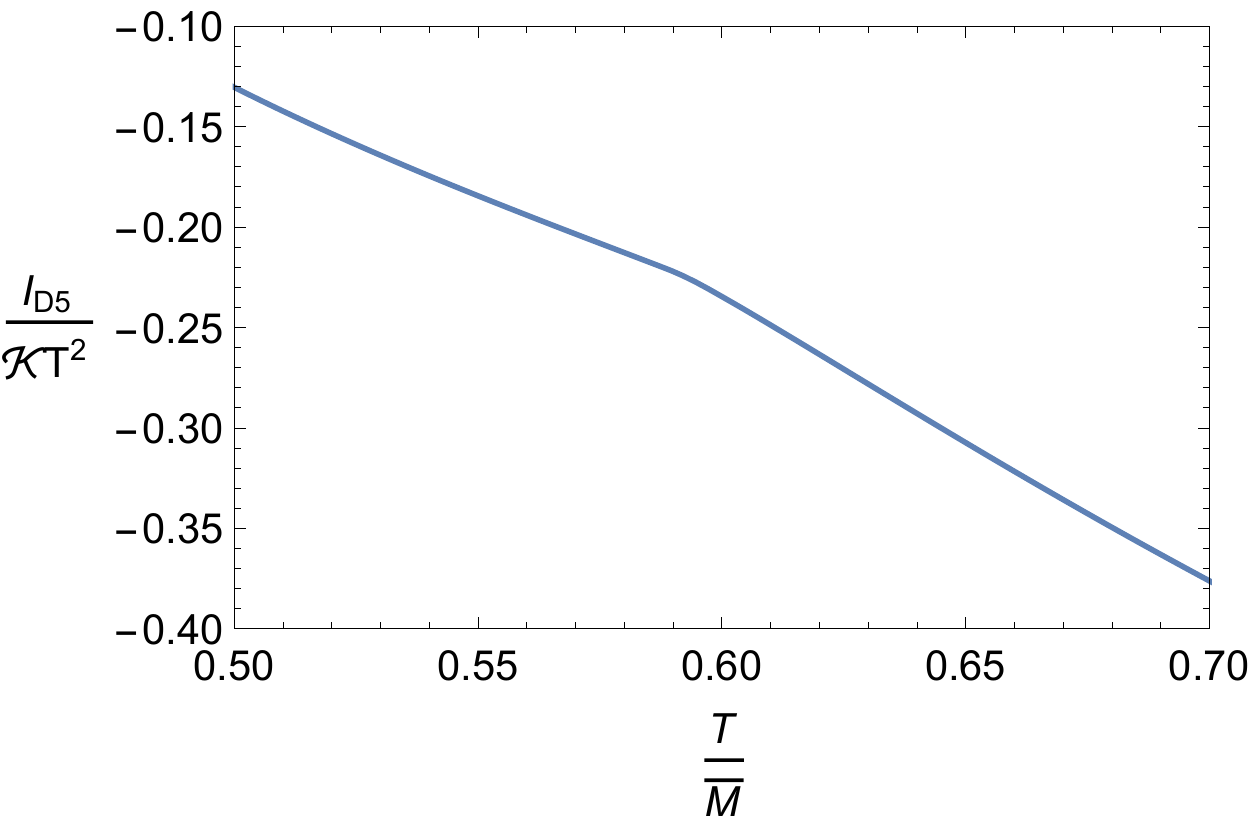}}
\subfigure[]{\includegraphics[scale=.65]{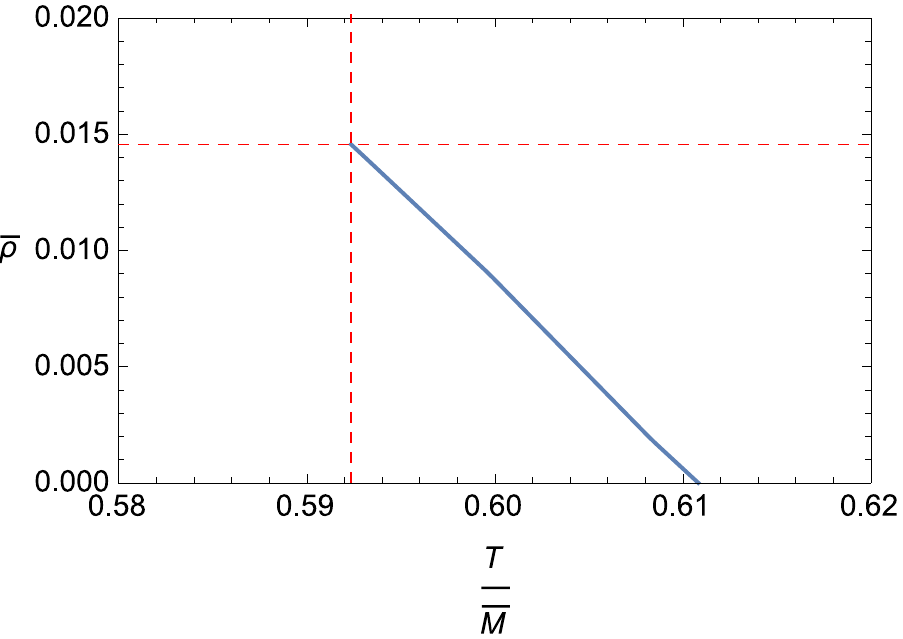}\quad}
\end{center}
\caption{(a) Regularised free energy of charged D5-branes with $\bho = 0.001, 0.1, 0.4, 0.8, 2, 4$ 
(from left to right). (b) Left panel: A close-up of the phase transition at low charge density 
($\bho=0.001$). Right panel: Phase transition is absent at high charge density ($\bho=0.015$).
(c) Charge density {\it vs.} critical temperature of phase transition at $\bho<\bho_*=0.0145$.}
\label{fig:FreeEQ}
\end{figure}

In order to decipher the phase diagram, we compare the on-shell free energy 
density \eqref{routhian} on different branches of solutions. The UV divergence encountered
as the D5-brane approaches the AdS boundary is regulated by introducing boundary 
counter-terms, as outlined in Appendix~\ref{app:thermodynamics-D5}. Numerical 
results for the regularised D5-brane free energy are 
shown in Figure~~\ref{fig:FreeEQ}(a) for different values of $\bho$.
At low charge densities, $\bho < \bho_*=0.0145$, we find a first order phase 
transition between two branches of black hole embedding solutions. The left- and 
right-hand panels in Figure~\ref{fig:FreeEQ}(b) showcase the different behaviour 
of the D5-brane free energy at $\bho<\bho_*$ and $\bho>\bho_*$, respectively.
Figure~\ref{fig:FreeEQ}(c) plots $\bho$ against the critical 
temperature of the phase transition and shows how the critical line
in the $T-\bho$ plane terminates at $\bho=\bho_*$

The low-$\bho$ phase transition connects to the phase transition between the 
black hole embedding and Minkowski embedding solutions at zero charge density. 
We note, in particular, that as $\bho\rightarrow 0$ the critical temperature of the 
phase transition between the different black hole embedding branches approaches 
$T_c/\bar M=0.611$, which is the critical temperature found in 
Section~\ref{sec:preliminaries} at zero charge density. 
Furthermore, the stable black hole embedding solution at low temperature 
and low charge density approaches a Minkowski embedding solution. 
It almost caps off at a finite radial distance outside the black hole, leaving a narrow throat 
that extends all the way to the horizon to accommodate the electric field lines emanating
from the black hole. 
In the $\bho\rightarrow 0$ limit the throat degenerates and the solution
takes the form of a Minkowski embedding solution. The onset of this low-temperature 
behaviour can be seen on the left in Figure~\ref{fig:psiVsrhoQ01} even if the D5-brane 
profiles in the figure are for a $\bho$ value somewhat above $\rho_*$.

A similar phase diagram, involving charged D7-branes in the finite temperature 
background of a black  3-brane, was worked out in \cite{Kobayashi:2006sb}. 
There it was argued that the favoured low-temperature configuration at charge 
densities below the analog of $\bho_*$ in the D7-brane system may in fact be 
unstable. In the present paper we are mainly concerned with evaluating two-point 
correlation functions of monopole operators in the presence of the probe D5-brane
and how they depend on the transverse spatial separation between the
monopole insertions on the boundary. As it turns out, we can determine this 
spatial dependence without having to rely on D5-brane solutions at very low $\bho$.
In what follows, we therefore restrict our attention to charged D5-branes with 
$\bho>\bho_*$, where there is only one branch of solutions and the question 
of an instability, analogous to the one discussed in \cite{Kobayashi:2006sb}, 
does not arise.

\subsection{Monopole two-point function at finite charge density}
\label{subsec:mz}

We now proceed to compute the action of a probe D3-brane ending on the charged 
D5-brane, which, under holographic duality, determines the two-point correlation function 
of monopole operators in a compressible finite charge density phase of the 
2+1-dimensional defect field theory~\cite{Iqbal:2014cga}.
The calculation is a straightforward generalisation from 
the charge-neutral case that was presented in Section~\ref{sec:monopole-nocharge}. 
The main new ingredient is the magnetic coupling between the probe D3-brane and 
the non-vanishing gauge field on the D5-brane world-volume. This means that the
second term in the D3-brane action \eqref{general_D3_action} in 
Appendix~\ref{app:convention} comes into play and the spacelike curve $\CC$ traced 
out by the probe D3-brane in the $\adsBH{4}$ part of the bulk geometry is no longer a 
geodesic in the rescaled metric~\eqref{gbar_mb}.
The endpoints at the $\ads_4$ boundary can still be taken to be at
$\up\rightarrow\infty$, $x\rightarrow \pm \frac{\Delta x}{2}$, $y\rightarrow 0$, but 
at intermediate points the curve extends in the $y$-direction and lies along 
$\{\up(\eta),x(\eta),y(\eta)\}$.
We refer the reader to Appendix~\ref{sec:D3_general} for the derivation of the
shape of $\CC$ and the regularised on-shell action of the probe D3-brane
at finite charge density. 
The main focus of the present section will instead be on presenting our numerical 
results and exploring the behaviour of the monopole equal-time two-point 
function as a function of spatial separation at different temperatures and 
charge densities. 

The Euclidean action of a D3-brane ending on a charged D5-brane is worked 
out in Appendix~\ref{sec:D3_general} below and is given by
\be\label{d3_action_final_main}
S_{D3}=
2N\int_{\up_*}^\infty \dif{\up}\,
\frac{\mu_b^2(\up)}{\sqrt{\up^2 \mu_b^2(\up)-2\bar{\mathcal P}^2\tilde f^{-1}(\up)}}
-\frac{\pi N\bar{\mathcal P}^2}{Q}\big(\omega\eta_e-\frac12\sinh{(2\omega\eta_e)}\big)\,.
\ee
where
\be\label{etai_exp_mainbody}
\omega \eta_e 
= \frac{2Q}{\pi}\int_{\up_*}^\infty \dif{\up}\,
\frac{1}{\up^2\tilde f(\up)\sqrt{\up^2 \mu_b^2(\up)-2\bar{\mathcal P}^2\tilde f^{-1}(\up)}} \,.
\ee
and $\eta\rightarrow\pm\eta_e$ at the endpoints of $\CC$ at the $\ads_4$ boundary. 
The frequency $\omega$, defined in \eqref{omegadef}, 
is the analog of a cyclotron frequency for a magnetic monopole in a 
background electric field. The curve has a turning point 
at $\eta=0$ at the radial coordinate $\up=\up_*$. The dimensionless constant of 
integration $\bar{\mathcal P}$ is a measure of transverse momentum in the $x$-direction.
It plays the same role as the parameter $\bar P$ in 
Section~\ref{sec:monopole-nocharge} and it is straightforward to see 
that $\bar{\mathcal P}\rightarrow \bar{P}$ as $Q\rightarrow 0$
(see Appendix C for details). In fact, the turning point analysis for a D3-brane 
ending on a D5-brane with a black hole embedding goes through unchanged,
with $\bar P$ replaced by $\bar{\mathcal P}$. 
When $\bar{\mathcal{P}} >\mubh$, 
the curve $\CC$ turns around at some $\upsilon = \upsilon_* > 1$ outside the horizon and 
returns to the boundary. On the other hand, for any 
$\bar{\mathcal{P}} \le \mubh$, the curve turns around at the horizon.

The first term in \eqref{d3_action_final_main} comes from the geometric DBI-action 
of the D3-brane and reduces to (\ref{Sd3-norho-mainbody}) for the uncharged case. 
The second term, which arises from the magnetic coupling between the probe D3-
and D5-branes, vanishes in the $Q\rightarrow 0$ limit. A regularised D3-brane action 
is obtained as before, by subtracting the action \eqref{sd3-disc-norho} of a 
$\bar{\mathcal P}=0$ curve. This cancels the divergence coming from the 
near boundary region $\up\rightarrow\infty$.

In order to determine how the monopole two-point function behaves as a function
of the endpoint separation, we plot the result of a numerical evaluation of the regularised
action against $\Delta x$ for different values of $\bar{\mathcal{P}}$ at fixed 
temperature and charge density. In Appendix~\ref{sec:D3_general} we obtain the
following expression for the endpoint separation in terms of dimensionless 
input parameters,
\be\label{def-DeltaxMbar-main}
\Delta x \bar{M}= { 2m \bar{\mathcal P}\over Q} \sinh \left(\omega  \eta _e\right)\,,
\ee
which can easily be evaluated numerically.
\begin{figure}
\begin{center}
\includegraphics[scale=0.83]{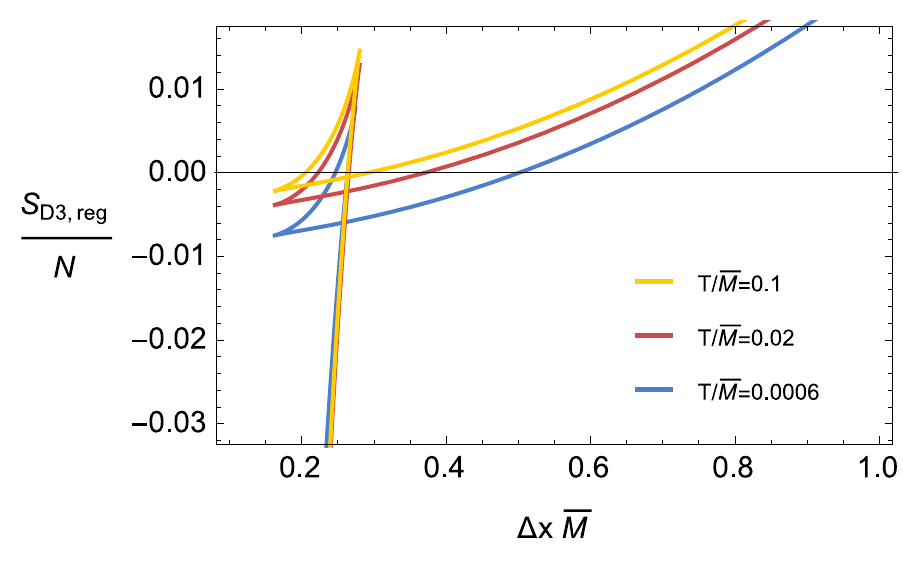}%
\includegraphics[scale=0.83]{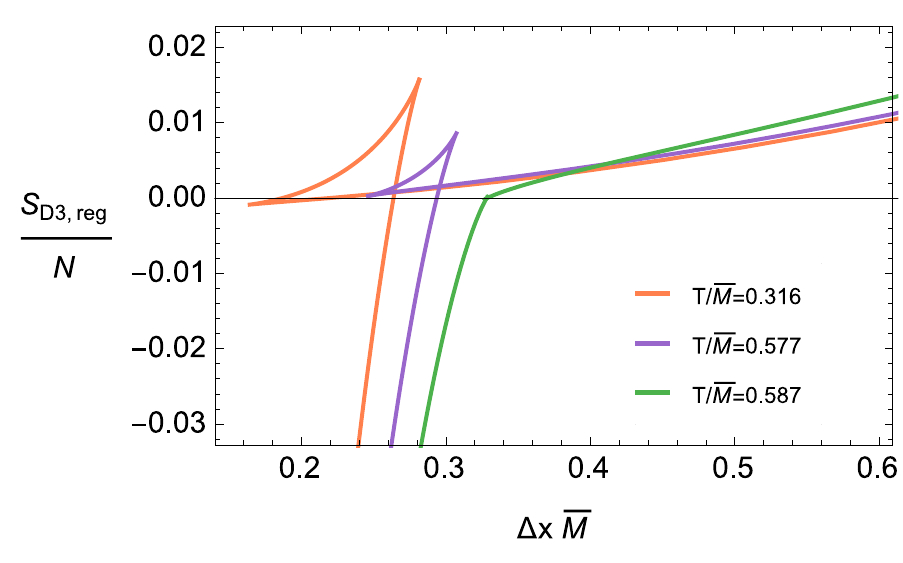}
\subfigure[$\bho = 0.02$]{
	\label{subfig:S3rho0p02}
\includegraphics[scale=.83]{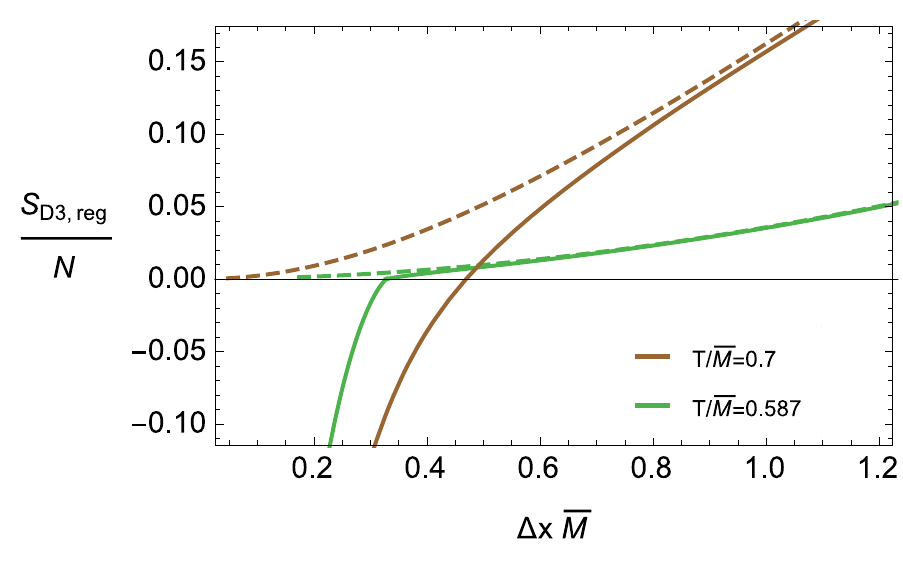}}
\subfigure[$\bho = 10$]{\label{subfig:S3rho10}
\includegraphics[scale=.83]{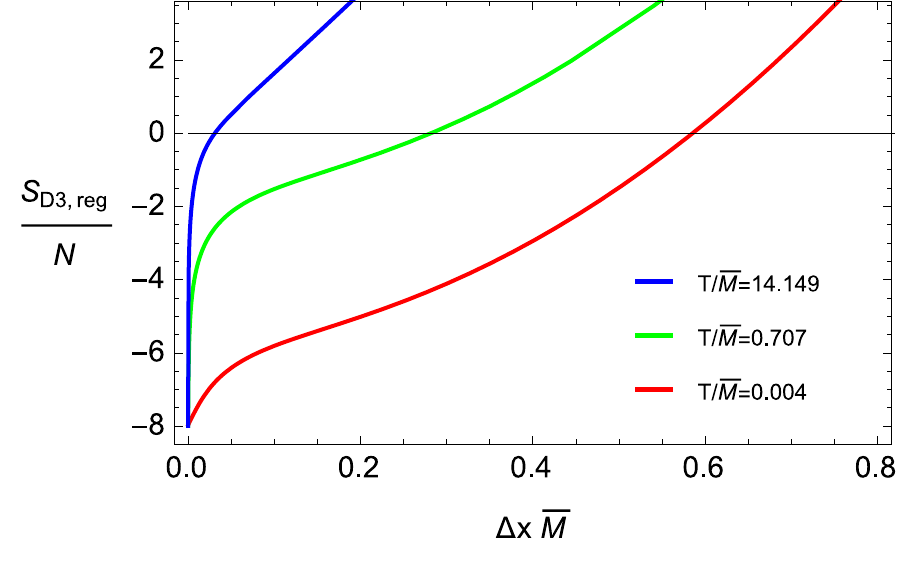}
	\includegraphics[scale=.83]{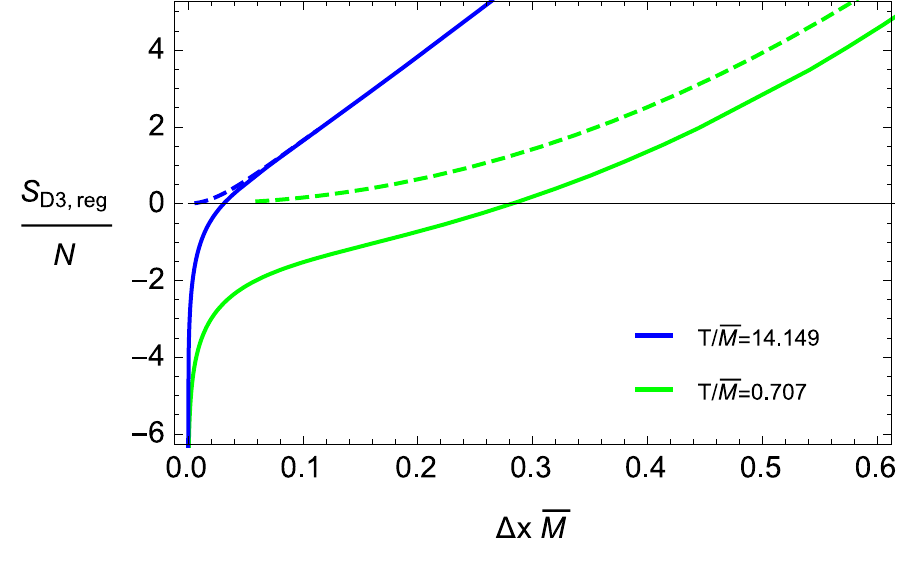}}
\end{center}
\caption{Euclidean D3-brane action against $\Delta x \bar{M}$ at (a) 
$\bho=0.02$ and (b) $\bho=10$. Each curve corresponds to a given temperature 
as indicated. 
The unstable $\bar{\mathcal{P}} <\mubh$ branch is indicated
by dashed lines in the middle and bottom right plots, but omitted from the others 
to avoid clutter.
}\label{fig:SD3vsDXQ10}
\end{figure}

\begin{figure}
\begin{center}
\includegraphics[scale=.78]{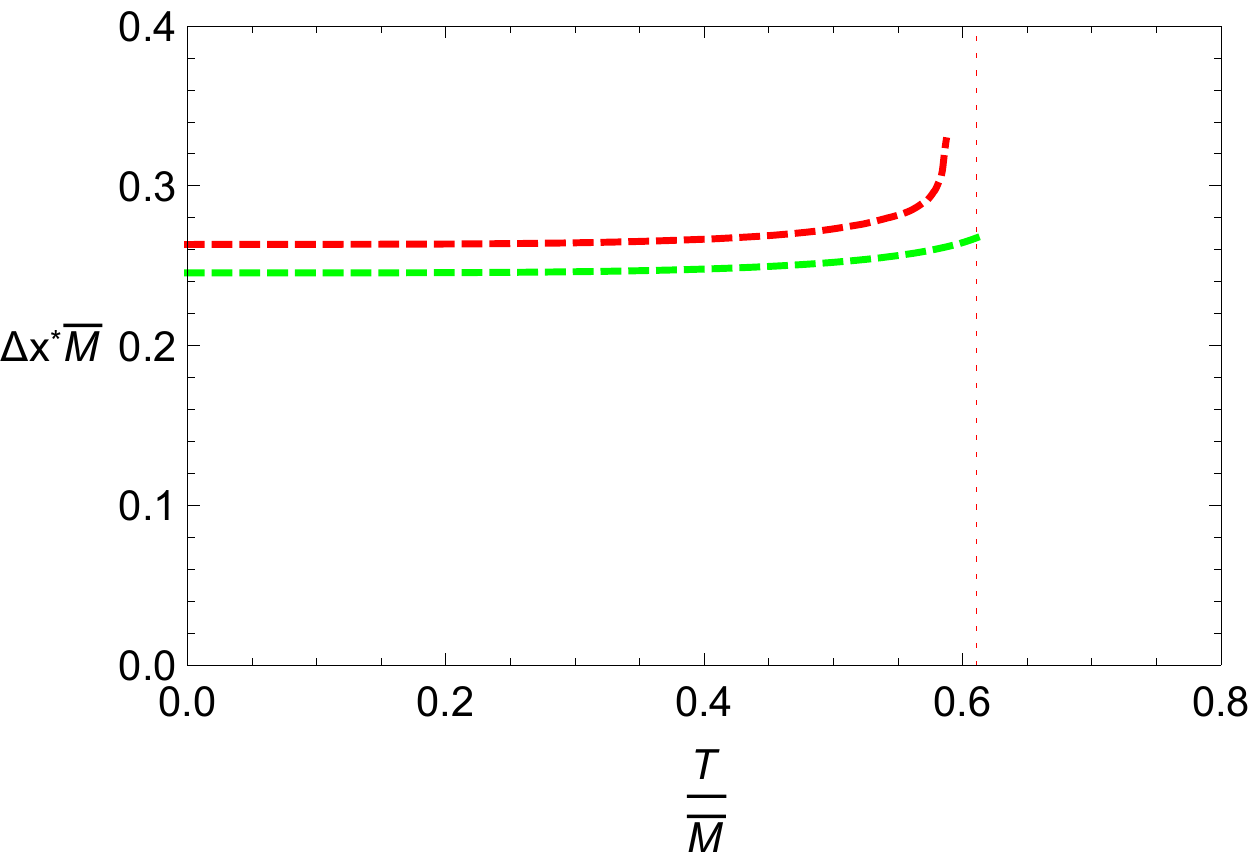}
\end{center}
\caption{The dashed red curve shows the critical endpoint separation $\Delta x^*$ 
for the first-order transition between two branches of solutions, at $\bho = 0.02$ as 
a function of temperature. The dashed green curve 
(repeated from \figref{DeltaXvsT} for reference) shows the endpoint separation at 
which the disconnected branch of D3-brane solutions becomes thermodynamically 
favoured in a Minkowski embedding at $\bho = 0$.
}
\label{fig:DxcT-rho02}
\end{figure}

The graphs in figure \ref{fig:SD3vsDXQ10} show our results for the regularised 
D3-brane action as a function of $\Delta x$ for several temperatures at two values
of the charge density: $\bho=0.02$ in Figure~\ref{fig:SD3vsDXQ10}(a) and 
$\bho=10$ in Figure~\ref{fig:SD3vsDXQ10}(b). 
At low charge density ($\bho\lesssim 0.1$) and low temperature
($T\lesssim 0.6 \bar M$ at $\bho=0.02$) we see evidence for a first-order
transition between two D3-branes that have 
different values of $\bar {\mathcal P}$. Both have 
$\bar {\mathcal P}>\mubh$ and thus the turning point is outside 
the horizon on both branches.\footnote{As in the charge-neutral case, 
there is also a branch with $\bar {\mathcal P}<\mubh$ and a turning point
on the horizon itself but this branch is never the most stable one. It is indicated 
with dashed lines in some of the graphs in Figure~\ref{fig:SD3vsDXQ10} but 
is left out of the others to avoid clutter.}
In the limit of vanishing charge density, this transition reduces to the transition 
between connected and disconnected D3-brane configurations that we saw for 
D5-branes in Minkowski embedding in the charge-neutral case in 
Section~\ref{sec:monopole-nocharge}. 
This is evident in Figure~\ref{fig:DxcT-rho02}, which plots the critical endpoint 
separation $\Delta x^*$ as a function of temperature at low charge density 
($\bho=0.02$) and compares it to that of the connected-disconnected transition. 

The graphs showing the regularised action of a connected D3-brane in 
Figure~\ref{fig:SD3vsDXQ10} share a common feature in that 
they all become convex at sufficiently large $\Delta x$. 
This can be traced to the magnetic term being 
dominant over the geometric DBI term in \eqref{d3_action_final_main}
at large endpoint separation while the DBI term governs 
the short-distance behaviour. A detailed analysis carried out in
Appendix~\ref{sec:highTmonopole} shows that the regularised 
D3-brane action always depends quadratically on $\Delta x$ in the 
``magnetic regime" at sufficiently large $\Delta x$. 
This, in turn, leads to Gaussian suppression of the equal-time
two-point correlation function of monopole operators as a function of the 
distance between the operator insertion points on the boundary.

The significance of the Gaussian suppression depends, however, 
on the value of the charge density in relation to the temperature. 
At high $\bho$ or low $T/\bar M$ the Gaussian behaviour sets in
at relatively short distances, as can for instance be seen in 
Figure~\ref{subfig:S3rho10}. This matches the zero temperature
results of \cite{Iqbal:2014cga} where the monopole two-point 
function was found to be Gaussian suppressed with
distance at finite charge density.
At low $\bho$ or high $T/\bar M$, on the other hand, the Gaussian behaviour 
only sets in at such long distances that the monopole two-point function is 
already vanishingly small due to the exponential suppression from
thermal screening discussed at the end of Section~\ref{sec:monopole-nocharge}. 

Our results for the monopole two-point function at finite $\bho$ and $T$ thus
interpolate nicely between Gaussian suppression obtained when the
charge density dominates over the temperature and exponential 
thermal screening found at low charge density.  

\section{Discussion}
\label{sec:discussions}

We have computed equal-time two-point correlation functions of magnetic monopole
operators in a strongly coupled 2+1-dimensional $\grU(1)$ gauge theory and studied
their spatial dependence. This provides information about the phase structure of the 
theory, including possible monopole condensation in a phase with a charge gap, and
also allows us to probe a compressible phase at finite $\grU(1)$ charge density.
Our investigation employs a top-down holographic construction involving intersecting 
D5- and D3-branes in $\ads_5\times \sphere^5$ spacetime, originally developed 
by Iqbal in~\cite{Iqbal:2014cga}, and extends it to 
an $\adsBH{5}\times \sphere^5$ black hole background in order to include thermal effects
and explore monopole correlators across the $\rho-T$ phase diagram.

In Section \ref{sec:zero-density} we focused on thermal effects on monopole 
correlators in a holographic phase with a charge gap at zero temperature.
The analysis performed in \cite{Iqbal:2014cga} showed that in this phase the 
holographic monopole two-point function at zero temperature saturates to a constant value 
as the separation between the operator insertions is increased. This is the expected 
behaviour when the monopole operator has condensed. On the gravitational side of 
the holographic duality the condensation is attributed to the vanishing of the bulk 
monopole effective mass where the D5-brane, on which the D3-brane representing
the bulk monopole ends, caps off in a Minkowski embedding. 
We find that the saturation of the monopole two-point function
at long distances persists at finite temperature, up to the critical temperature
where the D5-brane makes a transition from a Minkowski to a black hole embedding. 
Above the critical temperature, however, the monopole operator is no longer condensed
and the two-point function is exponentially
suppressed at long distances due to thermal screening.

In Section \ref{sec:finite_density} we turned our attention to thermal effects in a compressible
phase in the presence of a non-zero $\grU(1)$ charge density implemented by introducing a
gauge field on the D5-brane world-volume.  On the one hand, the finite charge density 
forces the D5-brane into a black hole embedding at any non-zero temperature and on 
the other hand it gives rise to a direct coupling between the D3-brane and the magnetic 
dual of the world-volume gauge field on the D5-brane. Studying the monopole correlation
function at finite temperature and charge density, we find that the transition to a 
disconnected D3-brane configuration at large separation found at vanishing charge
density is replaced by a transition
between different connected D3-brane configurations at low, but non-vanishing, charge 
densities and relatively low temperature. 

We also observe effects of the interplay at finite
temperature and charge density between the magnetic coupling and the geometric 
DBI-term in the D3-brane action. On the dual field theory side, this shows up in the
dependence of the monopole operator two-point function on spatial distance.
In particular, at finite charge density and low temperature the magnetic coupling 
contribution is the dominant one and the two-point function has a Gaussian falloff 
with distance. This is in line with the zero-temperature findings of \cite{Iqbal:2014cga}.
At high temperature and low charge density, on the other hand, the monopole two-point 
function exhibits exponential falloff due to thermal screening, but eventually crosses 
over to Gaussian suppression at very long distances. 

In this work we restricted our attention to the combined effects of finite charge density 
and temperature on two-point functions of monopole operators. An interesting and rather straightforward extension would be to add a non-zero magnetic field, for instance along 
the lines of \cite{Evans:2008nf, Filev:2009ai, Jensen:2010ga}.
Another future direction would be to relax some of the constraints that are built into the 
particular top-down holographic model used here in order to explore more general bulk 
monopole embeddings and the corresponding phase diagrams. 
Our treatment involved a probe D3-brane ending on a probe
D5-brane in an $\adsBH{5}\times \sphere^5$ background. An important question is
whether including the back-reaction of the D3/D5-brane system on the background geometry 
would stabilise or wash out the features we have found. Modelling the back-reaction may 
require the added flexibility of a bottom-up approach, while retaining essential features
of the top-down D-brane construction. At the same time, the study of monopole dynamics 
in phenomenologically motivated bottom-up models would be of considerable interest 
in its own right. 
Finally, there are other intersecting D-brane systems which can be used to model 
monopole operators in strongly coupled gauge theory. A D5/D7-brane model might for instance
be a more natural setting to study non-Abelian monopole correlators. 

\section*{Acknowledgements}

We thank N.~Evans, G.~Grignani, N.~Iqbal, N.~Jokela, M.~Lippert, R.~Myers, A.~Peet,
S.~Ross, and T.~Zingg for helpful discussions.
We thank J. \'A. Thorbjarnarson and P. R. Bryde for useful comments.
R.P. thanks the Perimeter Institute for 
Theoretical Physics for hospitality during the completion of this work.
This research was supported in part by the Icelandic Research Fund under contracts 
163419-051 and 163422-051, the Swedish Research Council under contract 
621-2014-5838, and by grants from the University of Iceland Research Fund.

\begin{appendix}

\section{Action functionals for probe D-branes}
\label{app:convention}

In this Appendix we collect some formulae and expressions which are  
used in the main body of the paper and in later appendices. We work out
the explicit form of the D5-brane action in the coordinate system used in
the main text. This is standard material but is included here in
order to have a self-contained presentation. 
We also obtain an explicit expression for the probe D3-brane action of a 
bulk monopole proposed by Iqbal in \cite{Iqbal:2014cga} in a black 3-brane
background. 

The tension of a D$p$-brane is given by
\be\label{brane_tension}
T_p= {1\over (2\pi)^p  g_s\, (\ell_s)^{p+1}}\,, 
\ee
where $g_s$ is the string coupling constant, and $\ell_s$ the string length,
which in turn are related to the 't Hooft coupling constant $\lambda$, the AdS radius $L$, 
and the number of background D3-branes $N$ as follows \cite{Maldacena:1997re},
 \be
\ell_s^2= \alpha'\,, \qquad g_s = {\lambda\over 4\pi N }\,, \qquad 
{L\over \ell_s} = \lambda^{1\over 4}\,.
\label{braneparameters}
\ee
 
\subsection{D5-brane}

The world-volume action for the probe D5-brane is 
 \be \label{SD5_full}
 S_{D5}= T_{5} \int_{D5} 2\pi\alpha' F  \wedge C_4-T_5 \int_{D5} \dif{^6\sigma}\,
 \sqrt{-\det\left(\gamma_{D5} 
 +2 \pi \alpha' F \right)}\,.  
 \ee
 where $C_4$ denotes the Ramond-Ramond 4-form field sourced by the background
 D3-branes, $F=dA$ the 2-form field strength of the D5-brane world-volume gauge field, 
 and $\gamma_{D5}$ the induced D5 world-volume metric. The first term in \eqref{SD5_full}
vanishes for the D5-brane configuration investigated in this work. To see this, we choose
a gauge where the 4-form $C_4$ is a sum of two terms, one proportional to the volume
form on $\R^4 \subset \adsBH{5}$ and the other to the volume form on the product of the two 
$\sphere^2$ factors inside $\sphere^5$. The former gives zero when wedged
with the 2-form $F$ while the latter has vanishing pullback to the
D5-brane world-volume.

In order to evaluate the remaining DBI term in \eqref{SD5_full} we take $\gamma_{D5}$
to be the induced metric in static gauge \eqref{gamma_5_norho} and parametrise the U(1)
gauge potential on the D5-brane world-volume as follows,
\be
A_t(\up)= {\sqrt\lambda \over 2\pi} a_t(\up)\,. 
\ee
After some straightforward algebra the D5-brane action reduces to 
\be
\label{D5_finitedensity}
S_{D5}= - {\sqrt{\lambda}\over 4\pi}  N V_3 T^2 \int \dif{\up} \left(1{-}\chi^2\right)\up^2 
\tilde f \sqrt{-\dot a_t^2+\frac{\pi^2\, T^2\,f^2}{2\tilde f}
\left(1+{{\up^2 \dot\chi}^2\over 1{-}\chi^2}\right)},
\ee
where $N$ is the number of background D3-branes, $V_3$ is the (infinite) volume from 
the integration over the $t,x,y$ variables, $T$ is the temperature of the 
background~\eqref{def_T}, and the dot denotes a derivative with respect to $\up$.
Since the action \eqref{D5_finitedensity} depends only on the derivative 
of the gauge potential  $a_t$, it is convenient to introduce a charge density,
\be\label{def_rho}
\rho=  {1\over V_3} {2\pi\over \sqrt\lambda }  {\delta S_{D5}\over \delta \dot a_t} 
= \half N \, T^2\, Q\,, 
\ee
with
\be\label{def_Q}
Q \equiv  { \dot a_t\,\up^2 \,\tilde f\, \left(1{-}\chi^2\right)\over \sqrt{-\dot a_t^2
+\frac{\pi^2\, T^2\,f^2}{2\tilde f}
\left(1+{{\up^2 \dot\chi}^2\over 1{-}\chi^2}\right)}}\,.
\ee
The equation of motion of the gauge field implies radial conservation of the 
charge density,
\be
\frac{d}{d\up}\,Q= 0\,.
\ee
We can take advantage of this by performing a Legendre transform on 
\eqref{D5_finitedensity} that trades the gauge 
potential $a_t$ for $Q$ as the independent field variable. This leads to an
action functional for $\chi$ that includes the conserved charge density $Q$ as a parameter,
\be
I_{D5}=
 \mathcal{K} \,T^2 \int \dif{\up}\, \up^2\, f\,
 \sqrt{\tilde f(1-\chi^2) \left(1-\chi^2+\up ^2 \dot{\chi}^2\right)} 
 \sqrt{\frac{Q^2}{\tilde f^2 \up ^4 \left(1{-}\chi^2\right)^2}+1}\,.
 \label{d5charged}
\ee
In order to study D5-brane thermodynamics, we have changed to Euclidean 
signature and taken Euclidean time to be periodic with period $1/T$. 
The temperature dependence of the constant in front of the action is left 
explicit and 
\be\label{kappadef}
\mathcal{K}\equiv \frac{\sqrt{\lambda} \,N \, V_2}{4\sqrt{2}}\,,
\ee
with $V_2$ the transverse area coming from the integral over $x$ and $y$.

The free energy of a D5-brane at charge density $Q$ is given by the on-shell 
value of the Euclidean action \eqref{d5charged}. The boundary counter-terms needed
to regularise the free energy are discussed in Appendix~\ref{app:thermodynamics-D5} and 
numerical results for the resulting regularised on-shell action are presented 
in Section~\ref{d5thermo}. Switching off the charge density $Q$ gives the free 
energy of a charge neutral D5-brane,
\be
I_{D5}=
 \mathcal{K} \,T^2 \int \dif{\up}\, \up^2\, f\,
 \sqrt{\tilde f(1-\chi^2) \left(1-\chi^2+\up ^2 \dot{\chi}^2\right)} \,,
 \label{d5neutral}
\ee
considered in Section~\eqref{sec:preliminaries}.

\subsection{D3-brane}

The action for the probe D3-brane is 
\be
\label{d3-dbi}
S_{D3}= T_3 \int_{D3} C_4
-T_3 \int_{D3} \dif{^4\sigma}\,\sqrt{-\det\left(\gamma_{D3} +2 \pi \alpha' F_{(3)} \right)}\,,
\ee
where $F_{(3)}$ is the field strength of the D3-brane world-volume gauge field and 
$\gamma_{D3}$  the induced D3-brane world-volume metric.
Gauge invariance of \eqref{SD5_full} and \eqref{d3-dbi} with respect to the $C_4$ gauge 
transformation,
\be
 \delta_\Lambda C_4= d \Lambda_3\,,
 \label{lambdatransf}
 \ee
requires the presence of additional terms,
\be\label{SK}
S_K= \int_{D5} K_3 \wedge dF+q_m \int_{\p D3} K_3\,,
\ee
involving a 3-form Lagrange multiplier $K_3$ that transforms as follows under the gauge transformation \eqref{lambdatransf}, 
\be
 \qquad \delta_\Lambda K_3 = -2 \pi \alpha' T_5 \Lambda_3 \,.
\ee
Note that this fixes the value of the coupling constant $q_m$ to be 
\be
q_m = {T_3 \over 2\pi\,\alpha'\, T_5}=  2\pi\,. 
\ee
The 3-form $K_3$ provides the magnetic coupling between the gauge field living 
on the D5-brane and the edge of the D3-brane. 
Indeed, adopting the same ansatz as in~\cite{Iqbal:2014cga},
\be\label{ansatz_K3}
K_3= {1\over 4\pi} \tilde A \wedge \omega_2\,, 
\ee
with $\omega_2$ the volume form on the two-dimensional unit sphere 
that the D5 and D3-branes wrap around inside $\sphere^5$, the 
three-dimensional term in \eqref{SK} reduces to an integral,
\be
q_m \int_{\mathcal C} \tilde A\,,
\ee
along the curve $\mathcal C$ in $\adsBH{4}$ traced out by the probe D3-brane.

The field strength of $\tilde A$ is the magnetic dual of the field strength of the 
D5-brane world-volume gauge field $A$ \cite{Iqbal:2014cga}. To see this, we vary the 
full action with respect to the field strength of $A$. 
Only two terms contribute, {\it i.e.} the DBI-term in \eqref{SD5_full} and the D5-brane
world-volume term in \eqref{SK}. Using the definitions \eqref{def_rho} 
and \eqref{ansatz_K3}, we obtain the following rather simple result,
\be
d\tilde A= \rho\, dx\wedge dy \,,
\ee
which is the magnetic dual of the radial electric field sourced by the
charge density $\rho$. It follows that we can choose a gauge where 
$\tilde A= -\rho\, y \,dx$. 

Finally, we collect the terms that contribute to the bulk monopole dynamics. 
Due to the specific D3-brane embedding employed in our analysis, the first term 
in \eqref{d3-dbi} containing the Ramond-Ramond potential $C_4$ vanishes and 
the remaining DBI term simplifies because there is no gauge field on the D3-brane
worldvolume. 
As explained in Section \ref{sec:monopole-nocharge}, for equal-time correlation 
functions the curve spanned in $\adsBH{4}$ by the probe D3-brane is spacelike 
and therefore the induced metric $\gamma_3$ has Euclidean signature. 
The relevant terms in the Euclidean D3-brane action are thus~\cite{Iqbal:2014cga} 
\be
\label{general_D3_action}
S_{D3}=T_3 \int_{D3}\dif{^4\sigma}\, \sqrt{\det \gamma_3} 
+i\, q_m \,\int_{\mathcal C} \tilde A\,.
\ee
In the charge-neutral case, considered in Section~\ref{sec:monopole-nocharge}, 
the second term is absent and the DBI term reduces to the action for a point particle
\eqref{dbi_particle} with a mass that depends on the radial position in $\adsBH{4}$.
In this case, the on-shell D3-brane action is simply given by the length of a 
geodesic in a rescaled metric where the position dependent mass has been
absorbed as a conformal factor, as discussed in Section~\ref{sec:monopole-nocharge}.
At finite charge density, on the other hand, the magnetic term in 
\eqref{general_D3_action} is non-vanishing and the curve $\mathcal{C}$ will no 
longer be a geodesic in the rescaled metric. This case is considered in detail in
Appendix~\ref{sec:D3_general}.

\section{Boundary counter-terms for D5-brane}
\label{app:thermodynamics-D5}

In the main text, we encountered several branches of D5-brane solutions. When
two or more different solutions exist for the same values of physical parameters it
is important to identify which solution is thermodynamically stable. For this, we
need to evaluate the free energy given by the on-shell Euclidean action of the D5-brane
and compare between different branches of solutions. In the charge-neutral case
this involves a comparison between Minkowski and black hole embedding solutions,
while at finite charge density we compare the free energy of different branches 
of black hole embedding solutions. 

As it stands, the D5-brane free energy \eqref{d5charged} is UV-divergent and needs
to be regularised by introducing appropriate counter-terms at the $\ads_4$ boundary. 
We use a well-established regularisation procedure for general D$p$-D$q$ systems 
described by DBI actions \cite{Karch:2005ms} and specialise to the system at hand. 
We take the UV cut-off surface to be at constant radial coordinate $\up=\up_{UV}$ and
find that the following counter-term action will cancel the UV-divergence of
the bulk D5-brane action,
\be\label{counter-terms}
S_{\rm b} = - \frac{\sqrt{\lambda}N}{6\pi^3 L^3} \int \dif{^3\xi}
\sqrt{\gamma^{\rm b}} 
\left( 1-\frac{3}{2}\chi^2+\ldots\right)\bigg|_{\up=\up_{UV}}\,,
\ee
where $\xi^i=(\tau,x,y)$ are boundary coordinates and $\gamma^{\rm b}_{ij}$ the 
induced metric at $\up=\up_{UV}$. 
A finite free energy is obtained by cutting off the integral at $\up=\up_{UV}$ 
in \eqref{d5charged}, or in \eqref{d5neutral} in the charge-neutral case, and 
evaluating the sum 
\be\label{reg_freeenergy}
I_{D5, \rm reg} = I_{D5}+S_{\rm b}
\ee
before taking the $\up_{UV}\rightarrow \infty$ limit. The ellipsis in \eqref{counter-terms}
denotes sub-leading terms that give a vanishing contribution in the limit. We note that
the presence of a gauge field on the D5-brane world-volume does not require 
any additional boundary counter-terms compared to the charge-neutral case. 

The regularised free energy can now be calculated numerically as a function of temperature 
for both black hole and Minkowski embeddings.\footnote{For efficient numerical 
evaluation of the free energy of a Minkowski embedding solution, it is convenient to
change to the $(r\,, R)$ variables introduced in Section~\ref{sec:preliminaries}.}
Results are shown in Figure~\ref{D5_action} for the charge-neutral case and in 
Figure~\ref{fig:FreeEQ} for D5-branes at finite charge density.

\section{On-shell D3-brane action}
\label{sec:D3_general}

In this Appendix we generalise the discussion of monopole two-point functions in
Section~\ref{sec:monopole-nocharge} to finite charge density. We obtain integral
expressions for the on-shell D3-brane action and for the endpoint separation on the 
$\ads_4$ boundary, in terms of dimensionless parameters that characterise the
background charge density $\rho$ and the curve $\mathcal{C}$ in $\ads_4$ that 
connects the two endpoints. Results from the numerical evaluation of these 
expressions are presented and discussed in Section~\ref{subsec:mz} 
of the main text.

Our starting point is the Euclidean D3-brane action \eqref{general_D3_action}.
Using the D3-brane world-volume coordinates $(s,\up,\theta,\phi)$ that were 
introduced in Section~\ref{sec:monopole-nocharge}, the action
reduces to
\be\label{d3_action_case2_app}
S_{D3} = N \int_{\CC} \dif{s}
\sqrt{\tilde G_{xx} \big({\dot x(s)}^2+{\dot y(s)}^2\big)+\tilde G_{\up\up}\, {\dot \up(s)}^2}
-i \, q_m\, \rho\int_{\CC} \dif{s}\, y(s)\, \dot x(s) \,,
\ee
where $\tilde G_{IJ}$ are components of the rescaled metric \eqref{gbar_mb}
and a dot denotes a derivative with respect to $s$.
The following Noether charges are conserved along the curve $\mathcal{C}$,
\be
P_x  = \tilde G_{xx}\, \dot x -i \,\omega\,  y\,, 
\qquad 
P_y = \tilde G_{xx}\, \dot y +i \,\omega\,  x\,,  
\ee
where 
\be\label{omegadef}
\omega \equiv  \frac{q_m \, \rho}{N}=\pi T^2 Q
\ee 
is the analog of a cyclotron frequency for a magnetic monopole in an electric field, 
and we have simplified the expression for the charges by using the constraint
\be\label{constraint_case2}
\tilde G_{xx}\big({\dot y( s)}^2+{\dot x( s)}^2\big)+\tilde G_{\up\up}\, {\dot \up(s)}^2={1}\,,
\ee
that follows if $s$ is taken to be an affine parametrisation of $\CC$.

The change of variables
\be\label{change_sl}
\tilde G_{xx} {d\over d s}= {d\over d \eta}
\ee
allows us to re-express the charges as
\be\label{Dx&Dy}
P_x= x'(\eta)- i \omega\, y(\eta)\,,\quad\quad
P_y  = y'(\eta)+ i \omega\, x(\eta)\,,
\ee
where prime denotes a derivative with respect to the new parameter $\eta$. 
We want to solve this system of first order differential equations subject to
suitable boundary conditions.
Without loss of generality, we can assume that the midpoint of the curve 
$\{\up(\eta),x(\eta),y(\eta)\}$ is at $\eta=0$ and the endpoints at $\eta=\pm\eta_e$.
The following conditions, 
\be\label{xybc}
x(0)=0, \qquad x(\pm\eta_e)=\pm \frac{\Delta x}{2}, \qquad y(\pm\eta_e)=0,\qquad
\up(\pm\eta_e)\rightarrow\infty,
\ee
then ensure that the curve intersects the boundary 
at $x\rightarrow \pm \frac{\Delta x}{2}$ and $y\rightarrow 0$.
A simple solution of \eqref{Dx&Dy} satisfying these conditions is given by
\be
x(\eta)=\beta\sinh{(\omega\eta)}\,, \qquad 
y(\eta)=i\beta\big(\cosh{(\omega\eta_e)}-\cosh{(\omega\eta)}\big)\,
\ee
where we have used translation symmetry in the $x$-direction to set $P_y=0$ and 
the parameters $\beta$ and $\eta_e$ are determined by the remaining Noether charge and
the endpoint separation through
\be\label{pxdeltax}
P_x =\beta\omega \cosh{(\omega\eta_e)}\,,\qquad
\Delta x = 2\beta \sinh{(\omega\eta_e)}\,.
\ee
By using \eqref{change_sl}, we can write the constraint \eqref{constraint_case2} as
\be\label{newconstraint}
\tilde G^{xx}\omega^2\beta^2+(\tilde G^{xx})^2\tilde G_{\up\up}\, \up '(\eta)^2={1}\,,
\ee
or equivalently
\be
\Big(\frac{\dif\eta}{\dif\up}\Big)^2
=\frac{(\tilde G^{xx})^2\tilde G_{\up\up}}{1-\omega^2\beta^2 \tilde G^{xx}}\,.
\ee
The value of the dimensionless product $\omega\eta_e$ at the endpoint can then be obtained as
an integral over the radial variable,
\be\label{etaintegral}
\omega \eta_e 
= \frac{2Q}{\pi}\int_{\up_*}^\infty \dif{\up}\,
\frac{1}{\up^2\tilde f\sqrt{\up^2 \mu_b(\up)^2-2\bar{\mathcal P}^2\tilde f^{-1}(\up)}} \,,
\ee
where $\up_*\equiv \up(0)$ denotes the turning point of the curve $\CC$ and  
\be\label{pbardef}
\bar{\mathcal P}\equiv \frac{\omega\beta}{\pi T}
\ee
is a dimensionless combination of input parameters.
The D5-brane embedding enters through $\mu_b(\up)$, the dimensionless effective 
mass of the bulk monopole defined in \eqref{mexp_mb}.

Similarly, the DBI-term in the D3-brane action \eqref{d3_action_case2_app}
can be written as a radial integral while the magnetic term can be 
obtained in closed form in terms of the endpoint variable in \eqref{etaintegral}, 
\be\label{d3action_integral}
S_{D3} = 2N\int_{\up_*}^\infty \dif{\up}\,
\frac{\mu_b^2}{\sqrt{\up^2 \mu_b^2(\up)-2\bar{\mathcal P}^2\tilde f^{-1}(\up)}}
-\frac{\pi N\bar{\mathcal P}^2}{Q}\big(\omega\eta_e-\frac12\sinh{(2\omega\eta_e)}\big)\,.
\ee
We note that the integral in the DBI term is logarithmically divergent. As discussed in 
Section~\ref{sec:monopole-nocharge}, we regulate the divergence 
by introducing an upper cut-off at $\up=\up_{max}$ in the integral and subtracting, 
for each boundary insertion point, the action of a {\it vertical\/} curve with $P_x=P_y=0$,
\be
S_{D3}^0=N\int_{\up_*}^{\up_{max}} \dif{\up}\,\frac{\mu_b(\up)}{\up}\,.
\ee
The integral in \eqref{etaintegral} 
is finite as it stands and does not require any regularisation.

For given values of the parameters $m$ and $Q$, that characterise a D5-brane 
embedding at a particular temperature and charge density, we evaluate the
regularised D3-brane action numerically for different
values of the dimensionless parameter $\bar{\mathcal P}$. 
The graphs in Figure~\ref{fig:SD3vsDXQ10} are obtained by plotting the result 
against the dimensionless combination
\be\label{deltaxbarm}
\Delta x\, \bar M = \frac{2m\bar{\mathcal P}}{Q}\sinh{(\omega\eta_e)}\,,
\ee
evaluated at the same parameter values using \eqref{etaintegral}.

In Section~\ref{sec:monopole-nocharge} we considered the corresponding calculation at
vanishing charge density. The formulae we used there can be obtained by taking the 
$Q\rightarrow 0$ limit at fixed temperature in the formulae in this Appendix. The correct 
limit is obtained by letting $\omega\rightarrow 0$ and 
$\beta\rightarrow\infty$ in \eqref{pbardef} while keeping the dimensionless parameter 
$\bar{\mathcal P}$  fixed. 

By inspecting \eqref{pxdeltax}, we immediately see that $\bar{\mathcal P}\rightarrow \bar P$ 
in this limit, where $\bar P$ is the dimensionless parameter used in the geodesic calculation 
in Section~\ref{sec:monopole-nocharge}. If we then insert \eqref{etaintegral} 
into \eqref{deltaxbarm} and take the limit, we obtain
\be\label{deltaxnorho}
\Delta x\, \bar M \Big\vert_{Q=0} =\frac{4m\bar P}{\pi} 
\int_{\up_*}^\infty \dif{\up}\,
\frac{1}{\up^2\tilde f(\up)\sqrt{\up^2 \mu_b^2(\up)-2\bar P^2\tilde f^{-1}(\up)}} \,,
\ee
which is the same as \eqref{dx-norho-invariant} employed 
in Section~\ref{sec:monopole-nocharge}.

We obtain the D3-brane action at vanishing charge density in a similar fashion.
The magnetic term in \eqref{d3action_integral} vanishes in the 
$Q\rightarrow 0$ limit, as can easily be seen by carrying out a small $\omega$ expansion
inside the parenthesis,
\be
-\frac{\pi N \bar{\mathcal P}^2}{Q}
\big(\omega\eta_e-\frac12\sinh{(2\omega\eta_e)}\big)
=O(Q^2) \,,
\ee
and the remaining DBI-term reduces to 
\be
S_{D3} \Big\vert_{Q=0}= 2N\int_{\up_*}^\infty \dif{\up}\,
\frac{\mu_b^2(\up)}{\sqrt{\up^2 \mu_b^2(\up)-2\bar P^2\tilde f^{-1}(\up)}} \,,
\ee
which is the integral in \eqref{Sd3-norho-mainbody} in 
Section~\ref{sec:monopole-nocharge}. 

\section{Asymptotic limits}
\label{asymptotia}

While the main part of the calculations in this work are numerical, 
it is nonetheless useful to compare the results to analytic approximations 
in the appropriate limits.
This is especially true on limits where either the parameters or the 
observables become very large or very small, since it can be hard to 
predict a priori at what point this causes a breakdown of the numerics.

In what follows, we will carry out such approximations and find that our 
numerics indeed remains reliable for the whole range of parameters studied.
Along the way, we will confirm certain asymptotic behaviours that the numerical 
calculations already suggest.

\subsection{D5 limits}

For the D5 backgrounds, we can complement the numerics by calculating the 
free energy analytically at all corners of the $(T, \bho)$ -plane.
As a matter of fact, the low-temperature case has two distinct limits: one where 
we first take $T \to 0$ while holding $\bho$ constant, in which case $Q \to \infty$, 
and another, where $\bho$, or equivalently $Q$ goes to zero first, and only then 
do we take $T \to 0$.

\subsubsection{High $T$, zero $\bho$}
\label{sec:highTbackground}

Let us first set $\bho = 0$.
Now the high-temperature limit for the D5-embedding can be solved semi-analytically, 
as is done for the D7/D3 system \cite{Mateos:2007vn}.
This is easiest to do in the original coordinate $u/u_0$, however we present the analysis in the $\up$ coordinates for consistency. 
High temperature, {\it i.e.} ${T\over \bar M}\gg1$, means a very small boundary mass 
$m\ll 1$, cf. \eqref{invtemp}. 
Thus, we need to expand $\chi (\up)$ around $\chi=0$.
The resulting linearized equation of motion has the solution
\begin{equation}
	\chi(\up)= \chi_{\mathrm{as}}(\up) \equiv {1\over\upsilon \sqrt{\tilde f(\up)}} \left(\,
	_2F_1\left(\frac{1}{4},\frac{1}{2};\frac{3}{4};\frac{4}{\up^4\tilde f(\up)^2}\right)
	-\frac{4 \sqrt{2} \Gamma
		\left(\frac{3}{4}\right)^2 \,
		_2F_1\left(\frac{1}{2},\frac{3}{4};\frac{5}{4};\frac{4}{\up^4\tilde f(\up)^2}\right)}{\up 
		\sqrt{\tilde f(\up)} \Gamma \left(\frac{1}{4}\right)^2}\right)
	\label{eq:HighTD5}
\end{equation}
in terms of the hypergeometric function $_2F_1$.%
\footnote{Notice that in terms of the original coordinates $u$, the argument in the hypergeometric functions is simply ${u^4_0\over u^4}$.}
We fixed the boundary conditions by requiring regularity at the horizon and unit boundary mass \cite{Mateos:2007vn}, that is
\be
\chi_\mathrm{as}(\up) \sim {1\over \up} + \frac{c_\mathrm{as}}{\up^2} 
+ \dots \,, \qquad \up \to \infty\,. 
\ee
The coefficient of the ${1\over \up^2}$ -term at the boundary is 
$c_\mathrm{as} = -\frac{4 \sqrt{2} \Gamma \left(\frac{3}{4}\right)^2}{\Gamma\left(\frac{1}{4}\right)^2} 
= -0.64622\ldots$.
To have an asymptotic solution corresponding to an arbitrary mass we write 
$\chi= m \chi_\mathrm{as}$.
For this solution, $c = m c_\mathrm{as}$.

Expanding the regularised free energy \eqref{reg_freeenergy} for a BHE around $\chi=0$ we get
\begin{equation}
\begin{split}
	{I_{D5, \rm reg} \over \mathcal{K} T^2} &\cong \int_{\up_{min}}^\infty \dif{\up} 
	\left \{\up^2\,  f \sqrt{\tilde f}\left(1-\chi^2+\half \up^2 \dot{\chi}^2 \right)- \up^2
	+{m^2 \over 2}\right \}\\
	&+\up_{min}\left({m^2\over 2}- {\up_{min}^2\over 3}\right)+ m c
= m^2 \gamma + v_{min} \frac{m^2}{2} - v_{min}^3 \frac{\tilde{f}^{\frac{3}{2}}(v_{min})}{3} 
+ m c
\end{split}
\end{equation}
where $\gamma$ is the integral
\begin{equation}
 \gamma = \int_{\up_{min}}^\infty \dif{\up} \left\{ \up^2\,  
 f \sqrt{\tilde f} \left(-\chi_\mathrm{as}^2+\half \up^2 \dot{\chi}_\mathrm{as}^2 \right)
 +\half\right\}.
\end{equation}
The integral is finite and can be numerically evaluated with the result $\gamma = 0.4693\ldots$.

Inserting $\up_{min}=1$, $c = m c_\mathrm{as}$, and $m={\bar M\over T}$, we obtain 
\be\label{highT-regD5-norho-app}
{I_{D5, \rm reg} \over \mathcal{K} T^2} \cong  - {2\sqrt 2\over 3} + \left({\bar M\over T}\right)^2 \left( \gamma+\half +c_\mathrm{as}\right) +\dots = - {2\sqrt 2\over 3} +0.323 \left({\bar M\over T}\right)^2+\dots\,.~~~
\ee
At very high temperature, at leading order the rescaled  free energy  is constant, as can also 
be seen in Figure \ref{D5_action}.
Therefore we see that the leading term has a $T^2$ behaviour, 
as expected in a conformal field theory  \cite{LANDSMAN1987141}.

\subsubsection{Low $T$, zero $\bho$} 

We can also obtain the low-temperature limit of the D5-action at zero charge density.
The coordinates \eqref{bgmetricu} are well suited for this calculation. We fix the point $u_m$ 
where the brane caps off, and then simply expand the equation of motion and action with respect 
to the horizon position $u_0$ and around the Minkowski embedding solution $\chi(u) = u_m/u$.

At the leading order, the only contribution comes from the finite part of the counter-terms, and 
we get
\begin{equation}
	\frac{I_{D5,\rm reg}}{\mathcal{K} T^2} = -\frac{\pi T}{8 \bar{M}} + \order\left(\frac{T}{\bar{M}}\right)^5,
\end{equation}
which shows that the free energy of the Minkowski embedding phase indeed goes smoothly to its 
$T = 0$ value.

\subsubsection{High $T$, high $\bho$}

Let us then consider the limit where both $T$ and $\bho$ are large.
The relevant asymptotic limit is where $\bho$ is of the same order as $T^2/\bar{M}^2$, in 
which case $Q = 2 \frac{\bar{M}^2}{T^2}\bho$ is finite.

In order to compute the regularised free energy to leading order at large $T$, we can proceed 
as before, and use a weak field expansion, that is $\chi\to 0$ (and $m\to 0$).
Hence, the regularised free energy \eqref{reg_freeenergy} for high-$T$, arbitrary $Q$, 
and dropping all terms proportional to $m$,
becomes:
\begin{eqnarray}
\frac{I_{D5, \rm reg}}{ \mathcal{K} T^2} = \int_{1}^\infty \dif{\upsilon} \left\{ \frac{\left(\upsilon^4-1\right)\sqrt{\frac{Q^2}{\upsilon ^4+1}+\frac{1}{\upsilon^4}+1}}{\upsilon ^2}-\upsilon ^2\right\}.
\end{eqnarray}

\begin{figure}
	\begin{center}
		\includegraphics[scale=.6]{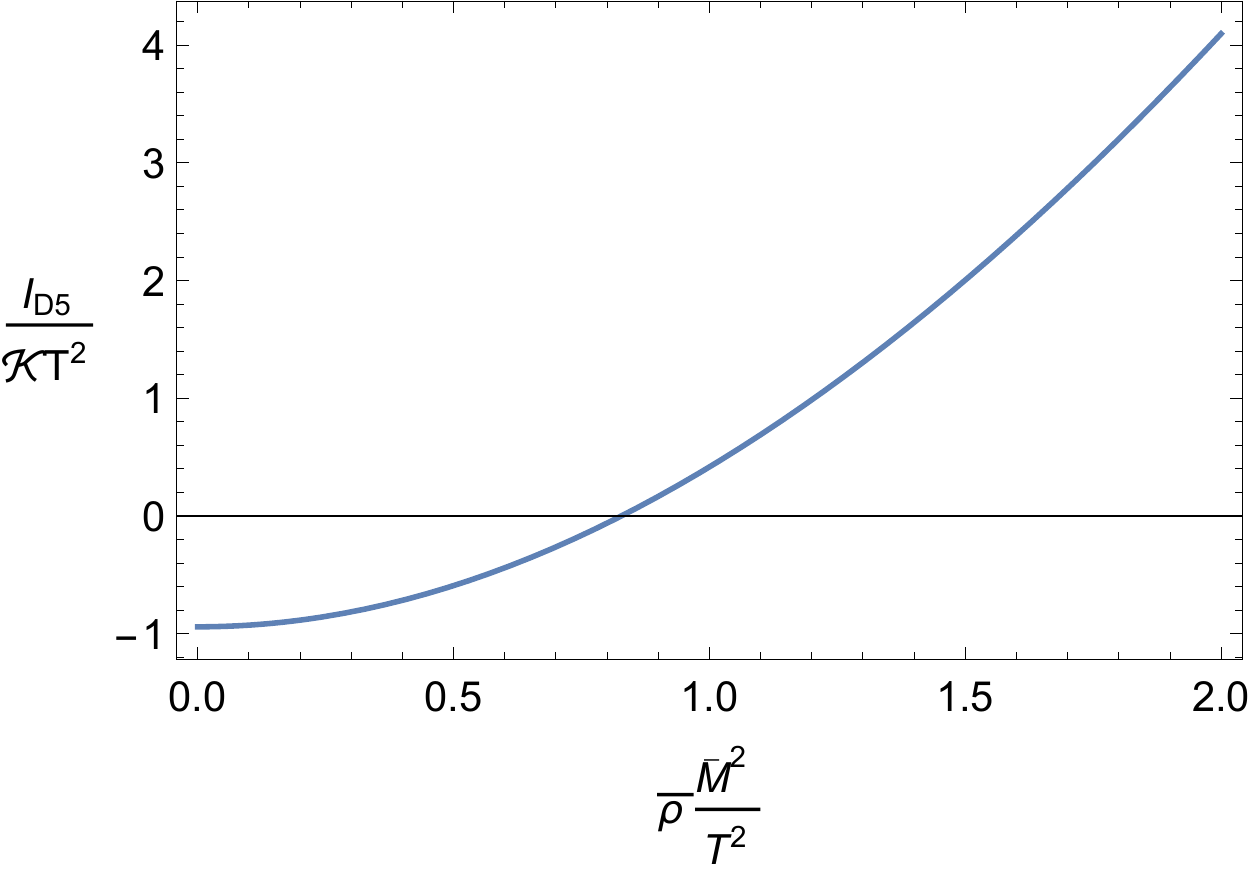}
	\end{center}
	\caption{The regularised free energy of the D5-brane at high-$T$ as a function of $\bho\frac{\bar{M}^2}{T^2}$.}\label{fig:R5highTQ}
\end{figure}

The integral can be evaluated explicitly using the change of variable $f(\upsilon)/\bar{f}(\upsilon) = \sin{y}$, with the result
\begin{equation}
\label{eq:D5highTcharged}
\begin{split}
\frac{I_{D5, \rm reg}}{ \mathcal{K} T^2} &=
\frac{1}{3} \left(-\sqrt{2} \sqrt{Q^2+4}-2 \sqrt{i Q} Q F\left(\left.i \sinh
^{-1}\left(\frac{\sqrt{i Q}}{\sqrt{2}}\right)\right|-1\right)\right) \\
&+ \order\left(\frac{\bar{M^2}}{T^2}\right),\quad T \gg \bar{M}
\end{split}
\end{equation}
where $F(\phi| k)$ is the elliptic integral of the first kind, with the convention that $k$, not $k^2$, appears in the integrand. The expression is real-valued, despite the presence of the imaginary 
unit, and it agrees with the numerical result in the high-$T$ limit.
\figref{fig:R5highTQ} exhibits the resulting free energy as a function of 
$\frac{1}{2}Q = \bho \frac{\bar{M}^2}{T^2}$.

Finally, in order to characterize the action as a function of the bulk charge, let us note that the asymptotic limits of \eqref{eq:D5highTcharged} are:
\begin{eqnarray}\label{highT-withQ}
\frac{I_{D5, \rm reg}}{ \mathcal{K} T^2} &=& -\frac{2 \sqrt{2}}{3}+\frac{Q^2}{2 \sqrt{2}}
+\order\left(Q^{5/2}\right),\quad Q \ll 1 \\
\frac{I_{D5, \rm reg}}{ \mathcal{K} T^2} &\propto& Q^{3/2} + \mathcal{O}(Q),
\quad Q \gg 1\,.
\end{eqnarray}
We note that \eqref{highT-withQ} matches the leading order 
of \eqref{highT-regD5-norho-app}, as expected. 

\subsubsection{Low $T$, finite $\bho$}

Finally, we consider the case where $T \to 0$ while $\bho$ remains finite, in which 
case $Q \to \infty$. Indeed, this limit needs to be considered since the free energy 
$I_{D5,\rm reg}$ diverges at low temperature.
The explanation is that the Lorentzian action changes by a finite amount when we 
go from zero temperature, zero charge, to finite charge, while still keeping temperature zero.
However, in $I_{D5,\rm reg}$, the Euclidean compactification introduces a factor of $1/T$, 
and therefore the free energy must indeed diverge as $1/T$ as $T \rightarrow 0$ at finite charge.

In terms of field configurations, at finite $\bho$, small $T$, the solution $\chi(\upsilon)$ remains 
nearly constant up to some finite $\upsilon$, and eventually breaks away to its asymptotic 
$m/\upsilon\  + \ldots$, \eqref{psi_boundary}, behaviour.
As temperature is further decreased, this breakaway point moves to larger and larger $\upsilon$. 

\begin{figure}
	\begin{center}
		\includegraphics[scale=.55]{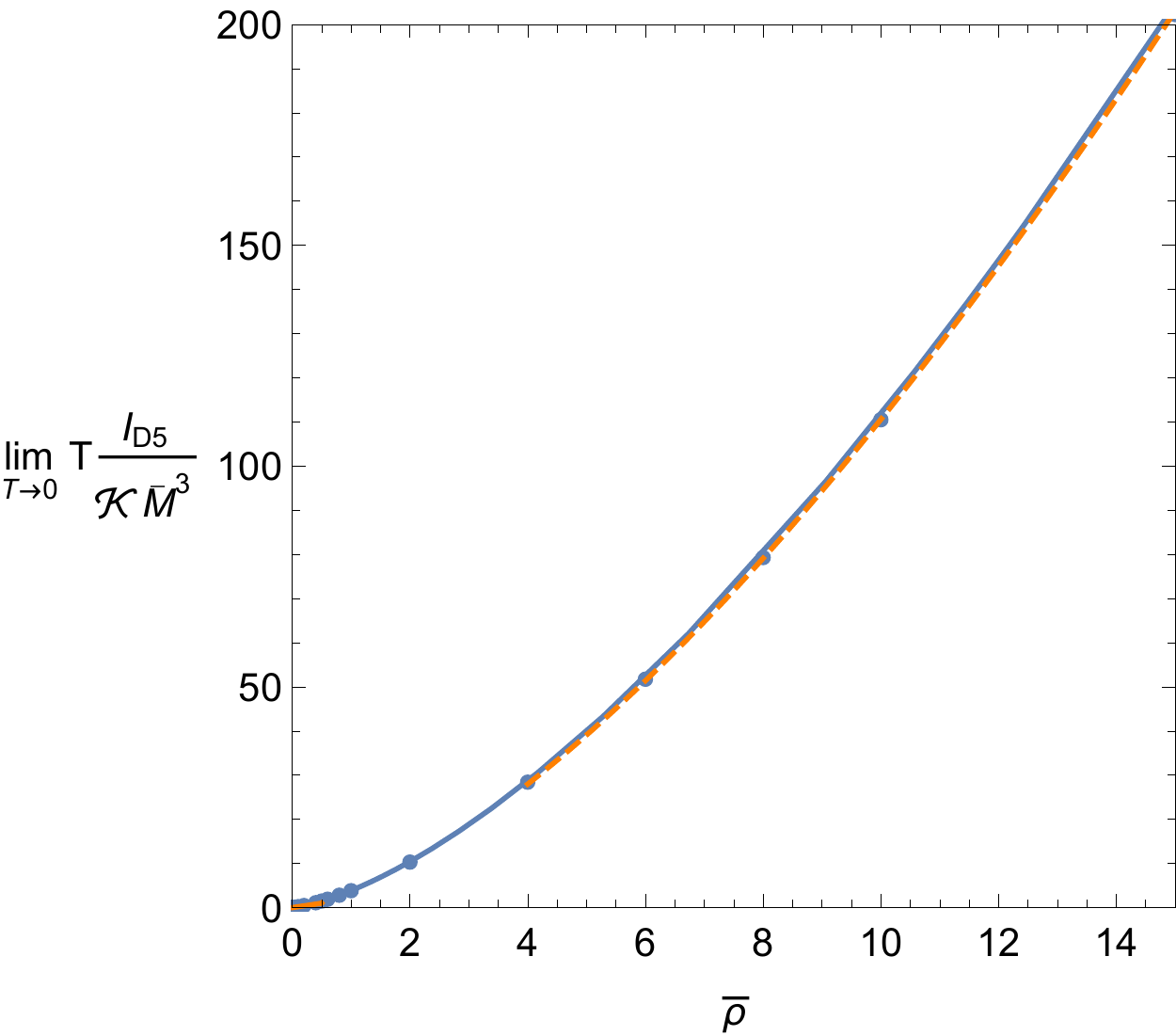}\quad\includegraphics[scale=.55]{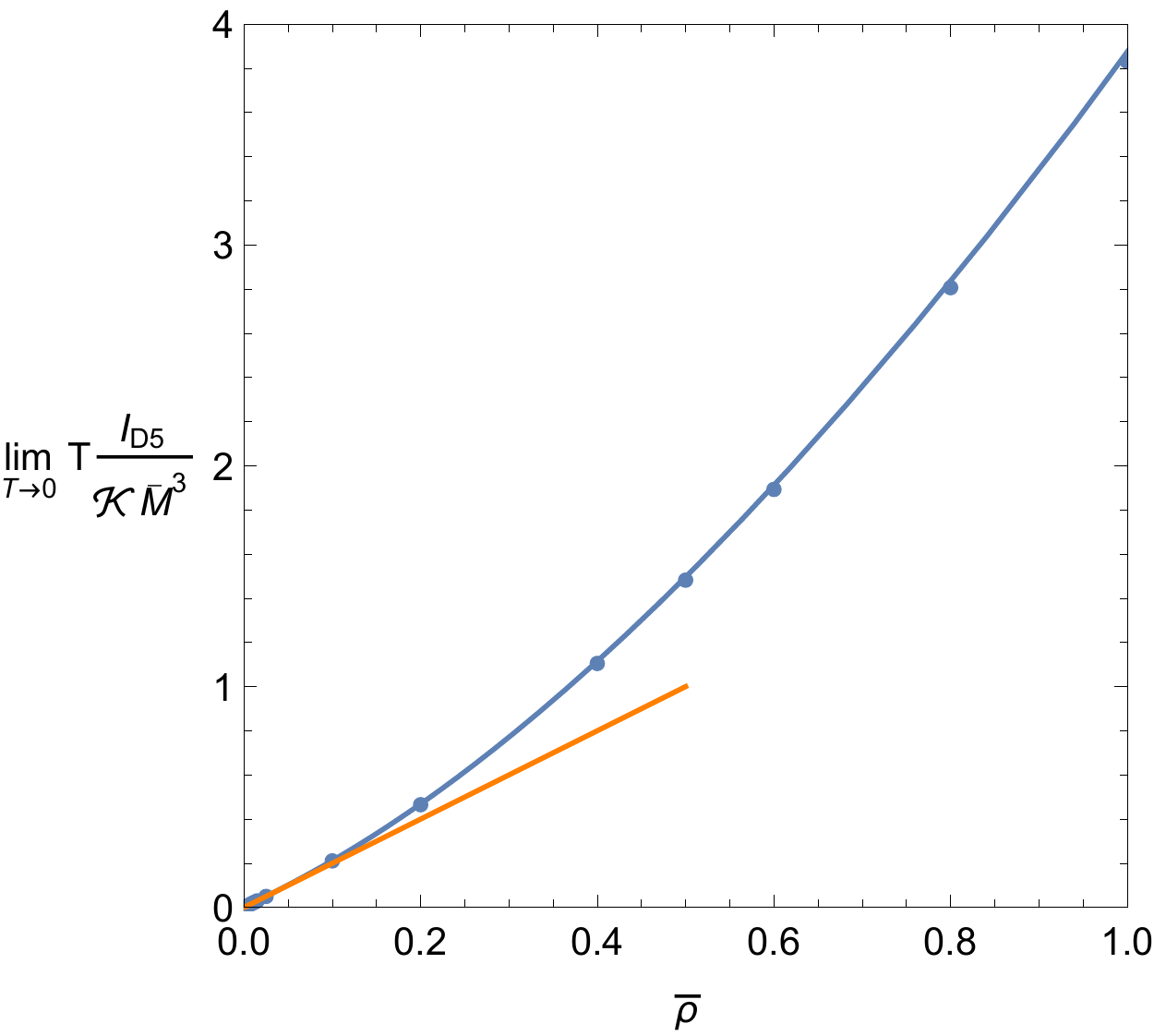}
	\end{center}
	\caption{Regularised free energy of charged D5-brane at zero temperature as a function 
	of the boundary charge.
		The points are evaluated from the smallest-T point available in the finite 
		temperature calculation, whereas the blue curve is from a calculation 
		where the zero-$T$ -limit is first 
		evaluated analytically, as described in the main text.
		The solid orange curve is the small-$\bho$ asymptotic, whereas the the 
		dashed orange curve in the left panel is the the large-$\bho$ asymptotic.
		Right panel shows a closeup of the small-$\bho$ -region.}
	\label{fig:zeroTfixedrho}
\end{figure}

We can explicitly compute $\lim_{T \rightarrow 0} T I_{D5,\rm reg}$, which is finite, by first 
changing variables to $w = u_0 (\upsilon - 1)$.
In these coordinates, the asymptotic behaviour appears at a finite value of $w$.
The resulting equations of motion have the large $w$ behaviour 
$\chi(w) = \hat{m}/w + \hat{c} / w^2 + \ldots$, and we see by comparison to 
\eqref{psi_boundary} that $\hat{m} = m u_0$, $\hat{c} = c u_0^2$.
Then $\rho = 1/2 N \bar{M}^2 \frac{\hat{Q}}{\hat{m}^2}$, where $\hat{Q} = Q u_0^2$.
Finally, we expand the free energy with respect to large $u_0$, keeping $\hat{Q}$ finite, 
solve numerically the resulting equations of motion with the boundary conditions 
$\chi(0) = \chi_0$, $\chi'(0) = 0$, and compute the value of $\bho$ and the free energy.
The boundary condition $\chi_0$ then controls $\bho$, whereas the bulk charge $\hat{Q}$ 
becomes simply a choice of scale, and its actual value is scaled out by a corresponding 
change in $\hat{m}$. Therefore we set $\hat{Q} = 1$ without loss of generality.
The resulting curve is shown in figure \ref{fig:zeroTfixedrho}.

Considering further the limit $\bho \rightarrow 0$, still at zero temperature, which 
corresponds to $\chi_0 \rightarrow 1$, we see that the solution tends to almost 
constant $\chi(w) = 1$ for small $w$, and then turns sharply to its asymptotic behaviour 
$\chi(w) \propto 1/w$ for large $w$.
Indeed, $\chi(w) = 1$ is an exact solution to the equation of motion, as is 
$\chi(w) = \hat{m}/w$ for any $\hat{m}$.
In order to move toward a global solution satisfying both of our boundary conditions, 
let us glue these solutions together at $w = w_s$.
In order to avoid a discontinuity, we must then choose $\hat{m} = w_s$.
This leaves a discontinuity in the derivative, with a jump of magnitude $1/w_s$.
Therefore, the glued function is a (weak) solution in the limit $w_s \rightarrow \infty$.
Keeping $\bho \propto \hat{Q} / w_s^2$ finite, we take that limit and evaluate the free energy, 
with the result
\begin{equation}
\lim_{T \rightarrow 0} \bar{M}^3 T \frac{ I_{D5,\rm reg}}{\mathcal{K}} 
= 2 \bho,\quad  \mathrm{when }\ \bho \ll 1.
\end{equation}

On the other hand, we can consider the limit $\bho \gg 1$ by expanding the regularised 
free energy and equations of motion around $\chi(w) \sim 0$.
The solution to the equation of motion to first order in $\chi_0$ is
\begin{equation}
\chi(w) = -\frac{(-1)^{1/4} \sqrt{\hat{Q}} \chi _0 F
\left(\left.\sin ^{-1}\left(\frac{(-1)^{3/4}w}{\sqrt{\hat{Q}}}\right)\right|-1\right)}{w},
\end{equation}
where $F(\phi | k)$ is as in \eqref{eq:D5highTcharged}, and again, the expression is real.
After evaluating the regularised free energy on this solution, we get
\begin{equation}
\lim_{T \rightarrow 0} \bar{M}^3 T \frac{ I_{D5,\rm reg}}{\mathcal{K}} 
= \frac{1}{3}\sqrt{\frac{2}{\pi }}  \Gamma^2 \left(\frac{1}{4}\right) \bho^{3/2},\quad  
\mathrm{when }\ \bho \gg 1.
\end{equation}

With these limits we now have analytic control over the D5 free energy in all four corners 
of the $(T, \bho)$ -plane, and we find that our numerical results match the approximations 
in all these cases.

\subsection{D3 limits}

\subsubsection{Large $\Delta x$ in the black hole embedding}
\label{sec:highTmonopole}

At the limit $\bar{\mathcal{P}} \rightarrow \mubh$ the integral in \eqref{etaintegral} and the 
DBI-term in \eqref{d3action_integral} diverge.
Therefore this gives the large -$\Delta x$ limit. The limit $\bar{\mathcal{P}} \rightarrow \mubh$ 
can be approached along two branches:
from below, where $\bar \CP<\mubh$, or from above, where $\bar\CP>\mubh$. 
In the first case ($\bar\CP<\mubh$) the turning point is at the horizon, and we will consider 
the following double limit: first $\up \to 1$ and then $\bar\CP \to \mubh$.
In the second case, $\bar\CP>\mubh$, the turning point is $v_*$, defined in \eqref{tp_location}, 
and we will consider the double limit $\up\to\up_*$ first, and then $\up_*\to 1 $, which is 
equivalent to $\bar\CP \to \mubh$. 

To illustrate the computations of the large $\Delta x$ limit in the BH embedding in a common 
set-up, we introduce $\up_t$, which takes value 1 in the $\bar\CP<\mubh$-branch, and $v_*$ 
in the $\bar\CP>\mubh$-branch. 
In both cases the dangerous integrand can be separated as
\begin{equation}
 l(\up;\Pbc) = \frac{1}{\sqrt{\up^2 \mu_b(\up)^2 - 2 \Pbc^2\tilde{f}^{-1}(\up)}},
\end{equation}
allowing us to write the DBI -term and the $\omega \eta_e$ integrals as
\begin{align}
 \omega \eta_e &= \frac{2 Q}{\pi} \int_{\up_t}^{\infty} \frac{l(\up;\Pbc)}{\up^2 \tilde{f}(\up)}\,,\\
 \frac{S_{D3,\mathrm{DBI}}}{N} &= 2 \int_{\up_t}^{\infty} \mu_b(\up)^2 l(\up;\Pbc).
\end{align}
We can further separate the divergent factor by writing
\be
l_1(\up;\bar\CP) \equiv {1\over\sqrt{\up^2 \mu^2_{b,t} - \frac{2\Pbc^2}{\tilde f(\up)}}}
\,, \qquad l_2(\up,\bar\CP)\equiv{l(\up;\bar\CP)\over l_1(\up;\bar\CP)}\,, 
\ee
where we have defined $\mu_{b,t}=\mu_{b}(v_t)$. In the limit we are considering 
$\mu_{b,t}$ is approaching $\mubh$. 
If we set $\Pbc = \mubh$, $l_1(\up;\Pbc)$ diverges as $1/(\up - 1)$, and therefore 
the integral diverges.
However, $l_2(\up; \Pbc)$ remains finite even in this case.
\footnote{Along the two branches the specific values of $l_2(\up;\bar\CP)$ are actually different.}

There is a slight difference in the form of $l_1(\up; \Pbc)$ between the two cases, 
$\Pbc < \mubh$ and $\Pbc > \mubh$, since in the latter case $\Pbc$ must be written in 
terms of $\up_*$ in order to capture the behavior of the function at $\up_*$.
However, in both cases an antiderivative of $l_1(\up; \Pbc)$ can be found exactly in 
terms of the incomplete elliptic integral of the first kind.
Denoting the antiderivative as
\begin{equation}
 L_1(\up;\Pbc) = \int \dif{\up}\ l_1(\up; \Pbc)
\end{equation}
for some fixed specific choice of the constant of integration, we can analyze its 
divergence at $\up = 1$ or $\up = \up_*$, respectively, as $\Pbc \rightarrow \mubh$.
We will find in both cases that $L_1(\up_t;\Pbc)$ diverges as $\log(|\Pbc - \mubh|)$.
We want to stress that the divergence in $L_1$ is characteristic of the limit 
$\bar\CP \to \mubh$, $\up_t\to 1$:
when $\Pbc > \mubh$ and therefore $\up_* > 1$, there is a one-over-square-root --divergence 
at the lower limit of the integral, which integrates to a finite number.
When $\Pbc < \mubh$, even the integrand itself is finite at $\up = 1$.

Now we write, for $\Pbc \neq \mubh$: 
\begin{equation}\label{int_l2}
\begin{split}
 \omega \eta_e &= \frac{2 Q}{\pi} \int_{\up_t}^{\infty} 
 \frac{l_2(\up;\Pbc)}{\up^2 \tilde{f}(\up)} l_1(\up;\Pbc) \\
 &= \left[\frac{2 Q}{\pi}\frac{l_2(\up;\Pbc)}{ \tilde{f}(\up) \up^2} 
 L_1(\up; \Pbc)\right]\Bigg|^\infty_{\up = \up_t} 
 - \frac{2Q}{\pi}\int_{\up_t}^{\infty}\frac{\dif{}}{\dif{\up}}
 \left(\frac{l_2(\up;\Pbc)}{\up^2\tilde{f}(\up)}\right)L_1(\up;\Pbc) \dif{\up}\,.
\end{split}
\end{equation}
The first term vanishes in the upper limit.
Integrating $L_1$ with respect to $\up$ gives a function of $\bar\CP$ that is bounded at 
$\up = \up_t$ for all $\Pbc \neq \mubh$, and the other factors in the integral are themselves 
bounded. Hence, in the limit $\bar\CP\to \mubh$ the second term in \eqref{int_l2} will 
give a finite result. Denoting such remaining integral term as 
$\frac{2Q}{\pi \tilde{f}(\up_t) \up_t^2} R_{\omega\eta_e}(\Pbc)$, we can solve 
\begin{equation}
l_2(\up_t;\Pbc) L_1(\up_t;\Pbc) = \frac{\pi \tilde{f}(\up_t) \up_t^2}{2 Q} \omega\eta_e + R_{\omega\eta_e}(\Pbc).%
\label{eq:divgfromomegaetae}
\end{equation}
We do the same integration-by-parts on $S_{D3,\mathrm{DBI}} - S_{D3}^0$ and use the above to write
\begin{equation}
\frac{S_{D3,\mathrm{DBI}} - S_{D3}^0}{N} = \mu_{b,t}^2 \frac{\pi \tilde{f}(\up_t)\up_t^2}{Q} \omega\eta_e + 2 \mu_{b,t}^2 R_{\omega\eta_e}(\Pbc) - R_{S_{D3}}(\Pbc).%
\label{eq:SD3fromomegaetae}
\end{equation}
We then insert this into \eqref{d3action_integral}, 
using \eqref{deltaxbarm} to express $\omega \eta_e$ in terms of $\Delta x$, to obtain the
regularised D3-brane action,
\begin{equation}
\label{finalD3action}
\frac{S_{D3,\rm reg}}{N} = 
(\tilde{f}(\up_t)\up_t^2\mu_{b,t}^2 {-} \Pbc^2) \frac{\pi}{Q}
\arcsinh\left(\frac{Q \Delta x \bar{M}}{2 m \Pbc}\right) 
{+} \frac{\pi \Pbc}{2 m} \Delta x \bar{M} \sqrt{1 {+}
 \frac{Q^2 \Delta x^2 \bar{M}^2}{4 m^2 \Pbc^2}} 
+ R(\Pbc),
\end{equation}
where
\begin{equation}
R(\Pbc) = 2 \mu_{b,t}^2 R_{\omega \eta_e}(\Pbc) - R_{S_{D3,\mathrm{DBI}}}(\Pbc)\,.
\end{equation}
This is still an exact result.
The key to deriving the large-$\Delta x$ limit is to now observe that since $L_1$ has a 
logarithmic divergence when $\Pbc \approx \mubh$, the first terms 
in \eqref{eq:divgfromomegaetae} and \eqref{eq:SD3fromomegaetae} 
dominate over the integral terms $R_{\omega\eta_e}$ and $R_{S_{D3,\mathrm{DBI}}}$, 
which are finite, as explained above.
Therefore, at leading order in the limit $\bar\CP\to\mubh$, the dependence on $\Pbc$ comes 
from the terms in \eqref{finalD3action} that contain $\Delta x$, and we obtain
\begin{equation}
\frac{S_{{D3,\rm reg}}}{N} \approx \mubh^2 \frac{\pi}{Q}\arcsinh\left(\frac{Q \Delta x 
\bar{M}}{2 m \mubh}\right) + \frac{\pi \mubh}{2 m} \Delta x \bar{M} \sqrt{1 + \frac{Q^2 
\Delta x^2 \bar{M}^2}{4 m^2 \mubh^2}}
+ R(\mubh)\,.
\label{eq:SD3largeDeltax}
\end{equation}
\begin{figure}
	\begin{center}
		\includegraphics[scale=.75]{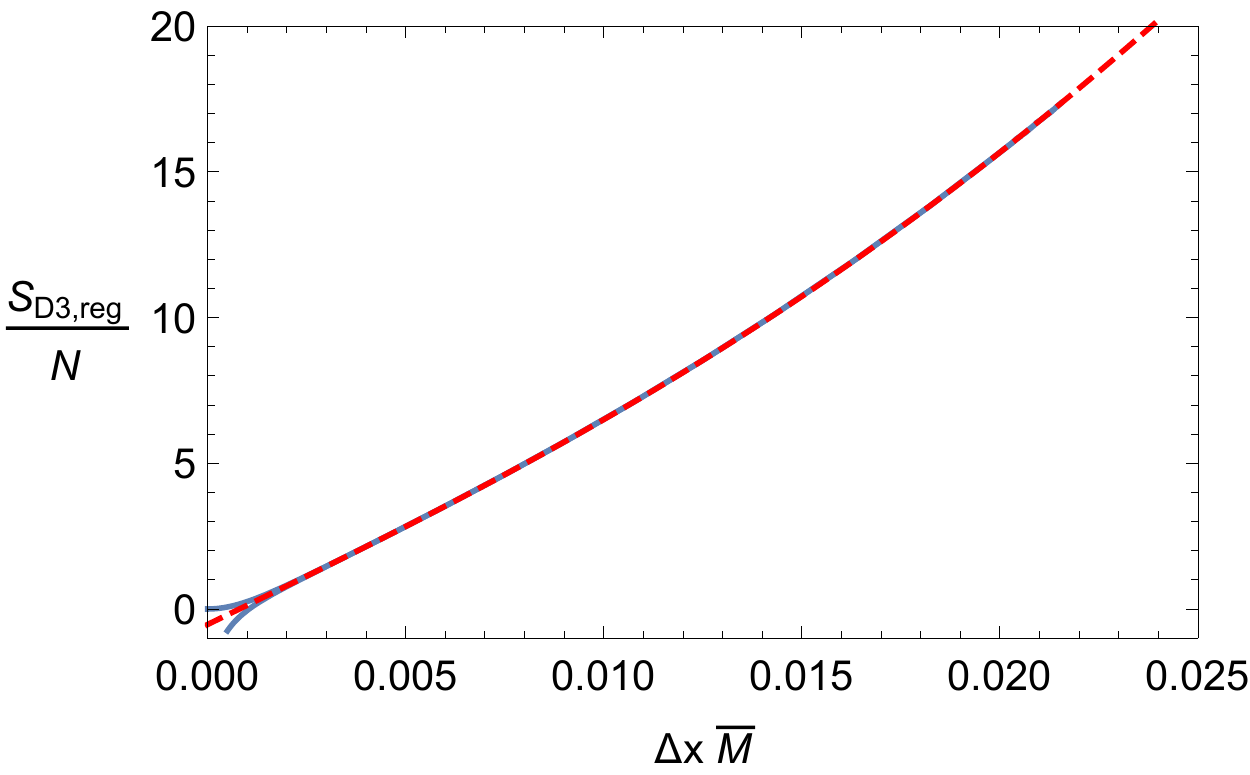} 
	\end{center}
	\caption{Comparison of our numerical results for the D3-brane action (blue curve) and 
	the approximation \eqref{eq:SD3largeDeltax} (dashed red curve) at $\bho \approx 13300$ 
	and $\frac{T}{\bar{M}} \approx 422$. These values of $\bho$ and $T$ were selected to 
	exhibit the crossover between linear and quadratic dependence on $\Delta x$. 
	The constant remainder term $R(\mubh)$ was fitted to match the largest $\Delta x$ 
	achieved by our numerics.
		}
	\label{fig:DeltaxAsymptote}
\end{figure}
When $\Delta x \bar{M} \ll \frac{m \mubh}{Q} $ we get
\begin{equation}
\frac{S_{{D3,\rm reg}}}{N} \approx \pi T \mubh \Delta x 
\xrightarrow{T \rightarrow \infty} \frac{\pi}{2} T \Delta x,
\label{eq:DeltaxlargeQsmall}
\end{equation}
and when $\Delta x \bar{M} \gg \frac{m \mubh}{Q}$
\begin{equation}
\frac{S_{{D3,\rm reg}}}{N} \approx \frac{\pi Q}{4 m^2} \Delta x^2 \bar{M}^2 
= \frac{\pi}{2} \bho \Delta x^2 \bar{M}^2.
\label{eq:DeltaxlargeQlarge}
\end{equation}
Note that when $Q$ is small but non-vanishing, there may be a long range of $\Delta x$ where 
the approximation \eqref{eq:DeltaxlargeQsmall} is valid before the system crosses over to its 
asymptotic behavior described by \eqref{eq:DeltaxlargeQlarge}.
At a given charge density and temperature, the crossover occurs at a distance 
where the two terms inside the square root in \eqref{eq:SD3largeDeltax} are equal, 
\begin{equation}
 \Delta x_{\mathrm{crossover}} \bar{M} = \frac{\mubh T}{\bar{M} \bho} 
 \xrightarrow{T \rightarrow \infty} = \frac{T}{2 \bar{M} \bho}.
\end{equation}
In other words, while the system always asymptotes to a quadratic behaviour at any finite 
charge, the crossover distance, where this behaviour takes over, diverges as $\bho$ 
goes to zero.

\subsubsection{Low-$\bar{\mathcal{P}}$ limit of the D3-brane action}
\label{app:low-P-limit}

At finite $\bho$ and low temperature, the argument of the $\sinh$-term in equations 
(\ref{d3action_integral}, \ref{deltaxbarm}) becomes large due to the factor $1/T^2$, and 
therefore the $\sinh$-term itself becomes extremely large.
This means that the most interesting region for $\Delta x$ is at extremely small 
values of $\bar{\mathcal{P}}$, to the extent that this region is not directly accessible 
to numerical calculations using standard tools.
In contrast, this region lends itself very well to an analytic low-$\bar{\mathcal{P}}$ approximation.
Specifically, let us take $\bar{\mathcal{P}} \ll \mubh$, and expand $\omega\, \eta_e$ to get
\begin{equation}
\label{eq:smallPappr1}
\omega \eta_e = \int_1^\infty \frac{4 \bho \bar{M}^2}{\pi \up^3  T^2\tilde{f} (\up)\mu_b(\up)}\dif{\up} 
+ \order\left(\frac{\bar{\mathcal{P}}^2}{\mubh^2}\right) = \frac{k}{T^2} 
+ \order\left(\frac{\bar{\mathcal{P}}^2}{\mubh^2}\right)
\end{equation}
where we have defined $k/T^2$ as the integral appearing in \eqref{eq:smallPappr1}.
Note that this is independent of $\bar{\mathcal{P}}$, and especially the $\sinh$-term is 
independent of $\bar{\mathcal{P}}$ in this approximation.
Then we get from \eqref{deltaxbarm} that
\begin{equation}
\label{eq:smallPappr2}
\bar{\mathcal{P}} = \frac{\bho \bar{M}^2}{T^2 \sinh(\frac{k}{T^2})}\Delta x.
\end{equation}
We then plug this into \eqref{d3action_integral}, observe that the geometric part is $S_{D3}^0/N + \frac{\pi}{2 \bho\bar{M}^2} \bar{\mathcal{P}}^2k$, and simplify to get
\begin{equation}
S_{D3, \rm reg} \equiv	S_{D3} - S_{D3}^0 
= \frac{\pi \bho N}{2} \frac{\Delta x^2 \bar{M}^2}{\sinh(\frac{k}{T^2})}\sqrt{1 
+ \sinh^2(\frac{k}{T^2})} + \order\left(\frac{\bar{\mathcal{P}}^2}{\mubh^2}\right).
\end{equation}
It is instructive to consider when this limit is valid in terms of $\Delta x$.
The ratio of the integrands in the leading and next-to-leading terms in \eqref{eq:smallPappr1} 
is $\bar{\mathcal{P}}^2/(\mu_b^2 \tilde{f} \up^2)$.
Approximating $\mu_b$ by $\mubh$, 
making the further approximation that the rest of the integrand can be treated as a constant, 
integrating, requiring that this ratio is small, and solving the corresponding $\Delta x$ from 
\eqref{eq:smallPappr2}, we get the approximate condition
\begin{equation}
\Delta x \bar{M} \ll \frac{T \sinh({\frac{k}{T^2})}\mubh}{\bho}.
\end{equation}
Comparing to numerics at not very low $T$, we see that indeed near this value 
of $\Delta x$ the approximation starts to deviate very significantly from the numerical results.
We note that in the region where $T$ is low and numerics becomes difficult 
due to the huge $\sinh$-term, the range of validity increases due to that very same term.
Therefore we can trust the approximation to give a very precise picture at low $T$ and 
moderate $\Delta x$.

Finally, we consider the low-$T$ limit,
\begin{equation}
S_{D3, \rm reg} = \frac{\pi}{2}\, \bho\,\bar{M}^2 N\Delta x^2\,,
\end{equation}
and, on the other hand, the zero-charge limit,
\begin{equation}
S_{D3, \rm reg} = \frac{\pi}{2}\,\frac{T^2}{\bar{k}}\,N\,\Delta x^2\,,
\end{equation}
where $\bar{k} = \frac{k}{\bho  \bar{M}^2}$.

\end{appendix}

\bibliographystyle{nb}
\bibliography{monopoles}

\end{document}